\documentclass[12pt]{article}

\usepackage{scicite}

\usepackage{times}
\usepackage{amssymb}
\usepackage{subfigure}
\usepackage{graphicx}
\usepackage[style=science,defernumbers=true,backend=bibtex]{biblatex}
\usepackage[final]{microtype}
\usepackage{csquotes,lmodern}
\usepackage{hyperref}

\hypersetup{hidelinks}

\def\nodata{.\,.\,.}

\def\aper{5\,kpc}
\def\FUVregularCut{25.15} %~mag~arcsec$^{-2}$}
\def\NUVregularCut{24.7} %~mag~arcsec$^{-2}$}
\def\SFRAREADENregularCut{$-2.25$}
\def\FUVtightCut{24.9} %~mag~arcsec$^{-2}$}
\def\NUVtightCut{24.45} %~mag~arcsec$^{-2}$}
\def\SFRAREADENtightCut{$-2.1$}

\def\redshiftrange{$0.02 < z < 0.09$}
\def\deltalow{$-0.35$}
\def\deltahigh{0.4}
\def\AVhigh{0.4}

\def\NUVMLCSeighteenHRtightCutsnnum{17}

\def\NUVMLCSeighteenHRtightCutall{0.073$\pm$0.012}

\def\NUVMLCSeighteenHRallsnnum{77}

\def\NUVMLCSeighteenHRallall{0.130$\pm$0.010}

\def\NUVMLCSeighteenHRregularCutsnnum{22}

\def\NUVMLCSeighteenHRregularCutall{0.083$\pm$0.012}

\def\NUVLOWERRMLCSeighteenHRtightCutsnnum{10}

\def\NUVLOWERRMLCSeighteenHRtightCutall{0.065$\pm$0.016}

\def\NUVLOWERRMLCSeighteenHRallsnnum{50}

\def\NUVLOWERRMLCSeighteenHRallall{0.128$\pm$0.014}

\def\NUVLOWERRMLCSeighteenHRregularCutsnnum{13}

\def\NUVLOWERRMLCSeighteenHRregularCutall{0.074$\pm$0.012}

\def\FUVMLCSeighteenHRtightCutsnnum{17}

\def\FUVMLCSeighteenHRtightCutall{0.074$\pm$0.012}

\def\FUVLOWERRMLCSeighteenHRtightCutsnnum{10}

\def\FUVLOWERRMLCSeighteenHRtightCutall{0.067$\pm$0.014}

\def\SFRAREADENMLCSeighteenHRtightCutsnnum{11}

\def\SFRAREADENMLCSeighteenHRtightCutall{0.075$\pm$0.018}

\def\SFRAREADENMLCSeighteenHRallsnnum{61}

\def\SFRAREADENMLCSeighteenHRallall{0.133$\pm$0.012}

\def\SFRAREADENMLCSeighteenHRregularCutsnnum{18}

\def\SFRAREADENMLCSeighteenHRregularCutall{0.094$\pm$0.015}

\def\NUVMLCSthirtyoneHRtightCutsnnum{17}

\def\NUVMLCSthirtyoneHRtightCutall{0.090$\pm$0.014}

\def\NUVMLCSthirtyoneHRallsnnum{78}

\def\NUVMLCSthirtyoneHRallall{0.138$\pm$0.008}

\def\NUVMLCSthirtyoneHRregularCutall{0.089$\pm$0.010}

\def\NUVLOWERRMLCSthirtyoneHRtightCutsnnum{8}

\def\NUVLOWERRMLCSthirtyoneHRtightCutall{0.094$\pm$0.027}

\def\NUVLOWERRMLCSthirtyoneHRallsnnum{46}

\def\NUVLOWERRMLCSthirtyoneHRallall{0.131$\pm$0.015}

\def\NUVLOWERRMLCSthirtyoneHRregularCutall{0.087$\pm$0.019}

\def\FUVMLCSthirtyoneHRregularCutsnnum{24}

\def\FUVLOWERRMLCSthirtyoneHRregularCutsnnum{13}

\def\SFRAREADENMLCSthirtyoneHRtightCutsnnum{11}

\def\SFRAREADENMLCSthirtyoneHRtightCutall{0.081$\pm$0.016}

\def\SFRAREADENMLCSthirtyoneHRallsnnum{62}

\def\SFRAREADENMLCSthirtyoneHRallall{0.139$\pm$0.012}

\def\SFRAREADENMLCSthirtyoneHRregularCutsnnum{18}

\def\SFRAREADENMLCSthirtyoneHRregularCutall{0.102$\pm$0.015}

\def\NUVMLCSeighteenHRtightCutmc{0.2\%}
\def\NUVMLCSeighteenHRregularCutmc{0.3\%}
\def\NUVLOWERRMLCSeighteenHRtightCutmc{1.1\%}
\def\NUVLOWERRMLCSeighteenHRregularCutmc{1.2\%}

\def\SFRAREADENMLCSeighteenHRtightCutmc{1.1\%}
\def\SFRAREADENMLCSeighteenHRregularCutmc{1.8\%}
\def\NUVMLCSthirtyoneHRtightCutmc{1.1\%}

\def\NUVLOWERRMLCSthirtyoneHRtightCutmc{24\%}

\def\FUVMLCSthirtyoneHRregularCutmc{0.6\%}

\def\FUVLOWERRMLCSthirtyoneHRregularCutmc{17\%}
\def\SFRAREADENMLCSthirtyoneHRtightCutmc{1.3\%}
\def\SFRAREADENMLCSthirtyoneHRregularCutmc{2.6\%}
\def\NUVMLCSeighteenHRtightCutpeculiartwohundred{0.050$\pm$0.010}

\def\NUVMLCSeighteenHRtightCutpeculiarthreehundred{0.075$\pm$0.015}

\def\rms{$\sim$\,0.065 to 0.075}

\def\smallmuerr{$\sigma_{\mu_{\rm SN}} < 0.075$\,mag}

\def\outlier{0.3\,mag}

\usepackage[labelfont=bf]{caption}

\renewcommand{\figurename}{Fig.}

\usepackage{filecontents}
\AtEveryBibitem{\clearfield{month}}
\AtEveryCitekey{\clearfield{month}}

\makeatletter

\csnumgdef{blx@entrycount}{0}
\AtEveryBibitem{%
  \csnumgdef{blx@entrycount}{\csuse{blx@entrycount}+1}}

\appto{\newrefsegment}{%
  \csnumgdef{blx@entrycount@\the\c@refsegment}{\csuse{blx@entrycount}+1}}

\defbibcheck{onlynew}{%
  \ifnumless{\thefield{labelnumber}}{\csuse{blx@entrycount@\the\c@refsegment}}
    {\skipentry}
    {}}

\makeatother

\newcommand\ionpat[2]{#1$\;${\scshape{#2}}}%                       % ion, i.e., CII = \ion{C}{ii}

\newenvironment{sciabstract}{%
\begin{quote} \bf}
{\end{quote}}

\renewcommand\refname{References and Notes}

\newcounter{lastnote}

\title{Distances with $<$4\% Precision from Type Ia Supernovae in Young Star-Forming Environments\footnote{This manuscript has been accepted for publication in Science. This version has not undergone final editing. Please refer to the complete version of record at http://www.sciencemag.org/. The manuscript may not be reproduced or used in any manner that does not fall within the fair use provisions of the Copyright Act without the prior, written permission of AAAS.}} %  with Twice Accuracy Possible with Phillips Relation}

\author{Patrick L. Kelly$^{1\ast}$, Alexei V. Filippenko$^1$, David L. Burke$^2$,\\
Malcolm Hicken$^3$, Mohan Ganeshalingam$^4$, WeiKang Zheng$^1$\\
\normalsize{Department of Astronomy, University of California, Berkeley, CA 94720-3411, USA.}\\
\normalsize{SLAC National Accelerator Laboratory, 2575 Sand Hill Road, Menlo Park, CA 94025, USA.}\\
\normalsize{Harvard-Smithsonian Center for Astrophysics, Cambridge, MA 02138, USA.}\\
\normalsize{Lawrence Berkeley National Laboratory, 1 Cyclotron Road, Berkeley CA 94720, USA.}\\
\normalsize{$^\ast$To whom correspondence should be addressed; E-mail:  pkelly@astro.berkeley.edu.}
}

\date{}

\begin{document}

\maketitle

\begin{sciabstract}
The luminosities of Type Ia supernovae (SNe), the thermonuclear explosions of white-dwarf stars, vary systematically with their intrinsic color and the rate at which they fade.  From images taken with the {\it Galaxy Evolution Explorer (GALEX)}, we identified SNe~Ia that erupted in environments that have high ultraviolet surface brightness and star-formation surface density. 
When we apply a steep model extinction law, we calibrate these SNe using their broadband optical light curves to %within \rms\,mag, a precision twice that measured for the parent sample of SN across all environments. 
within \rms~magnitudes, corresponding to $<$\,4\% in distance. 
The tight scatter, probably arising from a small dispersion among progenitor ages, suggests that variation in only one progenitor property primarily accounts for the relationship between their light-curve 
widths, colors, and luminosities.
\end{sciabstract}

\newrefsegment
The disruption of a white dwarf by a runaway thermonuclear reaction can yield a highly luminous supernova (SN) explosion.
The discovery that intrinsically brighter Type Ia supernovae (SNe) have light curves that fade more slowly \cite{ph93} and have bluer color \cite{ri96}
made it possible to determine the luminosity (intrinsic brightness) of individual SNe~Ia with an accuracy of $\sim0.14$ to 0.20\,magnitudes (mag) from only the SN color and the shape of the optical light curve. Through comparison between the intrinsic and apparent brightness of each SN~Ia, the distance to each explosion can be estimated. %Employing the 
With the precision afforded by light-curve calibration, measurements of distances to redshift $z \lesssim 0.8$ SNe~Ia showed that the cosmic expansion is accelerating \cite{re98,perlmutter99}.  Current instruments now regularly detect SNe~Ia that erupted when the universe was only $\lesssim4$\,billion years old \cite{jonesrodneyriess13}.

A growing number of large-scale observational efforts, including wide-field surveys capable of discovering large numbers of SNe~Ia, seek to identify the physical cause of the accelerating cosmic expansion \cite{weinbergmortonsoneisenstein13}.
Recent analyses have identified a $\sim10$\% average difference between the calibrated luminosities of SNe~Ia in low- and high-mass host galaxies \cite{kel10,sullivan10,lampeitl10,childressaldering13},
as well as a comparable difference between SNe~Ia with and without strong local H$\alpha$ emission within $\sim1$\,kpc \cite{rigaultcopin13}. %of the explosion site (uncorrected for extinction) may be $\sim0.1$\,mag less luminous, after light-curve correction \cite{rigaultcopin13}. 
Study of the intrinsic colors of SNe~Ia \cite{xwang09,foleykasen11} has additionally found that the colors depend on the expansion velocities of the ejecta near maximum light. %While models of the evolution with redshift in the population of SN and galaxies may be approximate,
With accurate models that include possible evolution with redshift, corrections for these effects should improve cosmological constraints from SN distances.

Sufficiently precise calibration of SNe~Ia would sharply limit the impact of these potential systematic effects, as well as others, on cosmological measurements, so recent efforts have sought to improve calibration accuracy by examining features of SN~Ia spectra \cite{baileyaldering09,xwang09,foleykasen11,blondin11,silvermanganeshalingam12,mandelfoleykirshner14}, infrared luminosities \cite{woodvaseyfreidman08,mandelnarayan11}, and their host-galaxy environments \cite{kel10,sullivan10,lampeitl10,rigaultcopin13}.
Flexible models for light curves, including principal component analysis, have also recently been applied to synthetic photometry of SN spectral series \cite{kimthomas13}. 
These several approaches that use additional data about the SN beyond its broadband optical light curve may enable calibration of luminosities to within $\sim0.10$ to 0.12\,mag ($\sim5$ to 6\% in distance). %, they require measurements significantly more costly to acquire than broadband optical imaging.

Here we show that a subset of SNe~Ia, identified only from photometry of a circular $r=$ \aper\ aperture at the SN position, yield distances from optical light-curve fitting with $<4$\% precision. Our sample consists of SNe~Ia with \redshiftrange\, and is assembled from the Lick Observatory Supernova Search (LOSS), Harvard-Smithsonian Center for Astrophysics (CfA), and Carnegie Supernova Project (CSP) collections of published light curves given in Table~\ref{tab:lcs}. Table~\ref{tab:sample} describes our light-curve sample selection criteria.   The \redshiftrange\ redshifts of the SNe place them
in the Hubble flow, where galaxy peculiar velocities are substantially smaller than velocities arising from the cosmic expansion.

We computed distance moduli using the MLCS2k2 light-curve fitting algorithm \cite{jha07} available as part of the SuperNova ANAlysis (SNANA; v10.35) \cite{kes09b} package. MLCS2k2 parameterizes light curves using a decline parameter $\Delta$ and an extinction to the explosion $A_V$, and solves simultaneously for both. Model light curves with higher values of $\Delta$ fade more quickly and are intrinsically redder.

To model extinction by dust, MLCS2k2 applies the O'Donnell \cite{odonnell94} law, parameterized by the ratio between the $V$-band extinction $A_V$ and the selective extinction $E(B-V) = A_B - A_V$.  Whereas this ratio $R_V = A_V/E(B-V)$ has a typical value of $\sim 3.1$ along Milky Way sight lines \cite{fitzpatrickmassa07}, lower values of $R_V$ yield the smallest Hubble-residual scatter for nearby SNe~Ia \cite{hi09b}.
A Hubble residual (HR $\equiv \mu_{\rm SN} - \mu_z$) is defined here as the difference between $\mu_{\rm SN}$, the distance modulus to the SN inferred from the light curve, and $\mu_z$, the distance modulus expected from the SN redshift and the best-fitting cosmological parameters.
SN~Ia color variation that does not correlate with brightness \cite{scolnicriess14}, or reddening by dust different from that along Milky-Way sight lines \cite{patattaubenberger14} may explain why low values of $R_V$ yield reduced Hubble-residual scatter. 
We find that the value $R_V = 1.8$ minimizes the scatter of Hubble residuals of the SNe in our full \redshiftrange~sample,
after fitting light curves with values of $1 \leq R_V \leq 4$ in $\Delta R_V=0.1$\,mag increments.

\begin{figure*}
\centering
\subfigure{\includegraphics[angle=0,width=4.5in]{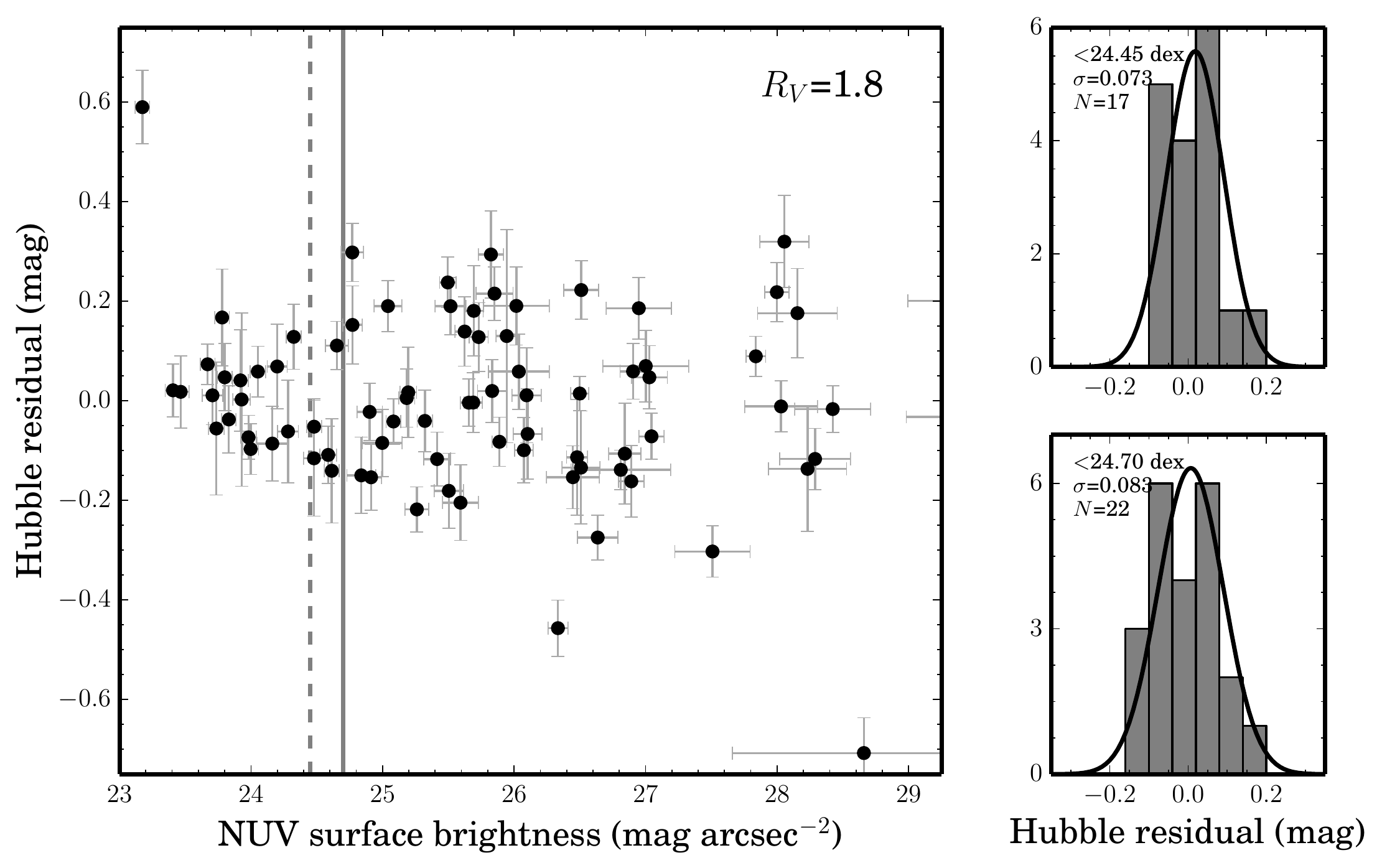}}
\subfigure{\includegraphics[angle=0,width=4.5in]{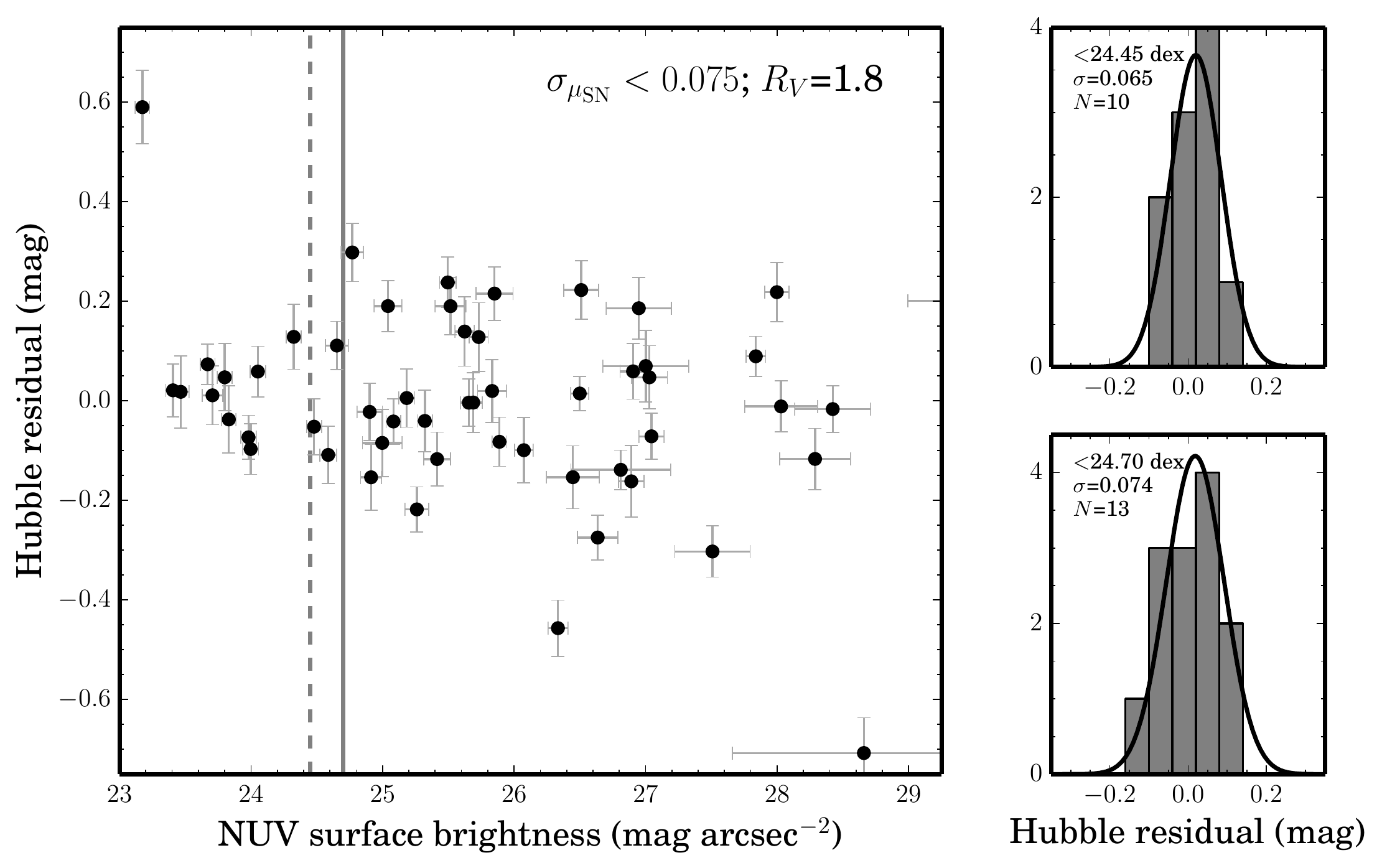}}
\caption{Hubble residuals against NUV surface brightness within the $r=$ \aper\ aperture around SNe~Ia. 
Top panels show all SNe in the sample, whereas bottom panels include only SNe with small statistical uncertainty (\smallmuerr) in 
distance modulus. Panels on the right show the distributions of Hubble residuals for SNe in regions brighter than the \NUVregularCut\,mag\,arcsec$^{-2}$ marked by a solid vertical line in each figure. Dashed vertical lines show a more restrictive threshold of \NUVtightCut\,mag\,arcsec$^{-2}$.  
Error bars shown above and in other plots correspond to 68\% confidence intervals. 
 }
\label{fig:NUV}
\end{figure*}

\begin{figure*}
\centering
\subfigure{\includegraphics[angle=0,width=6.5in]{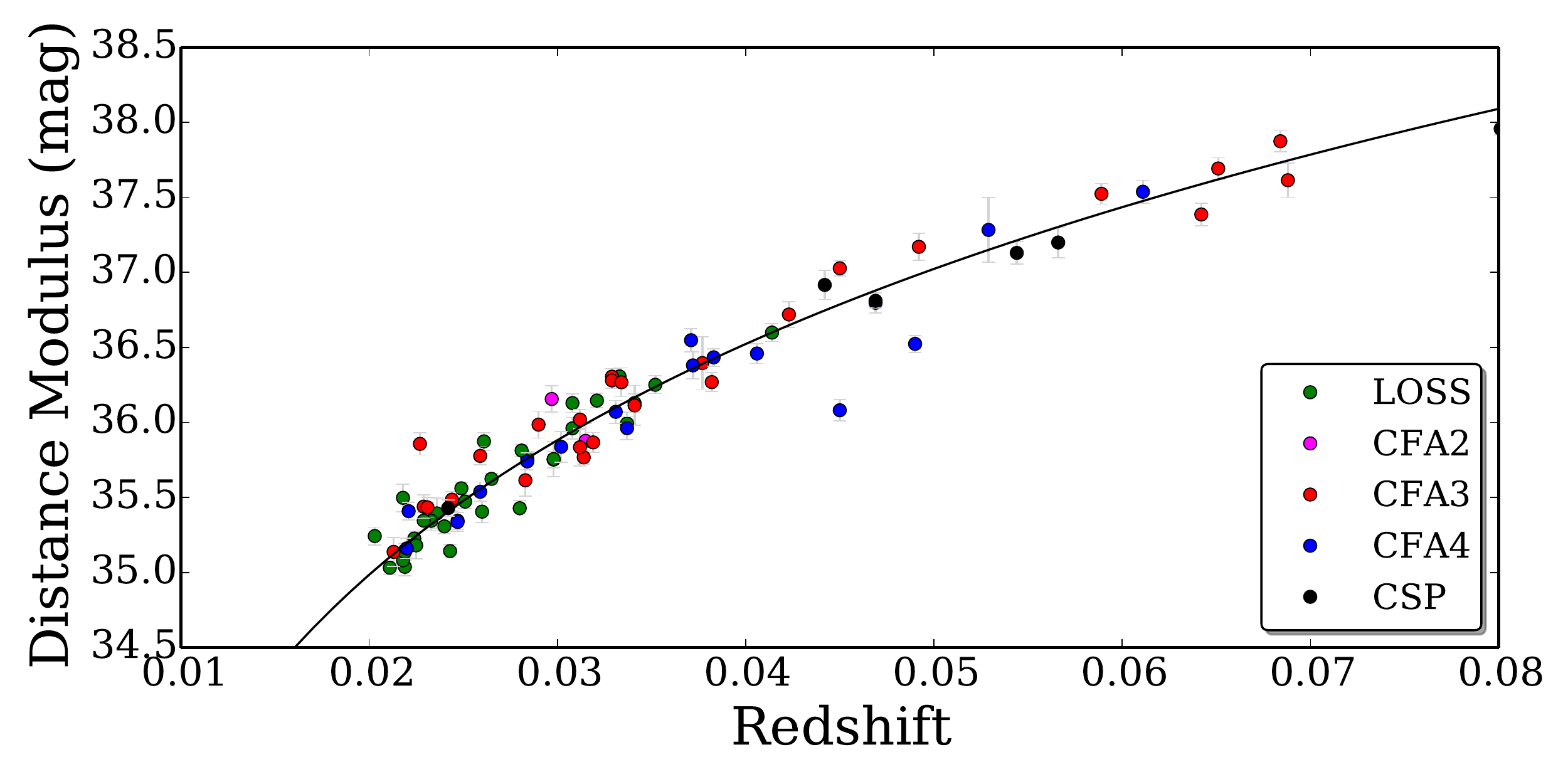}}
\subfigure{\includegraphics[angle=0,width=6.5in]{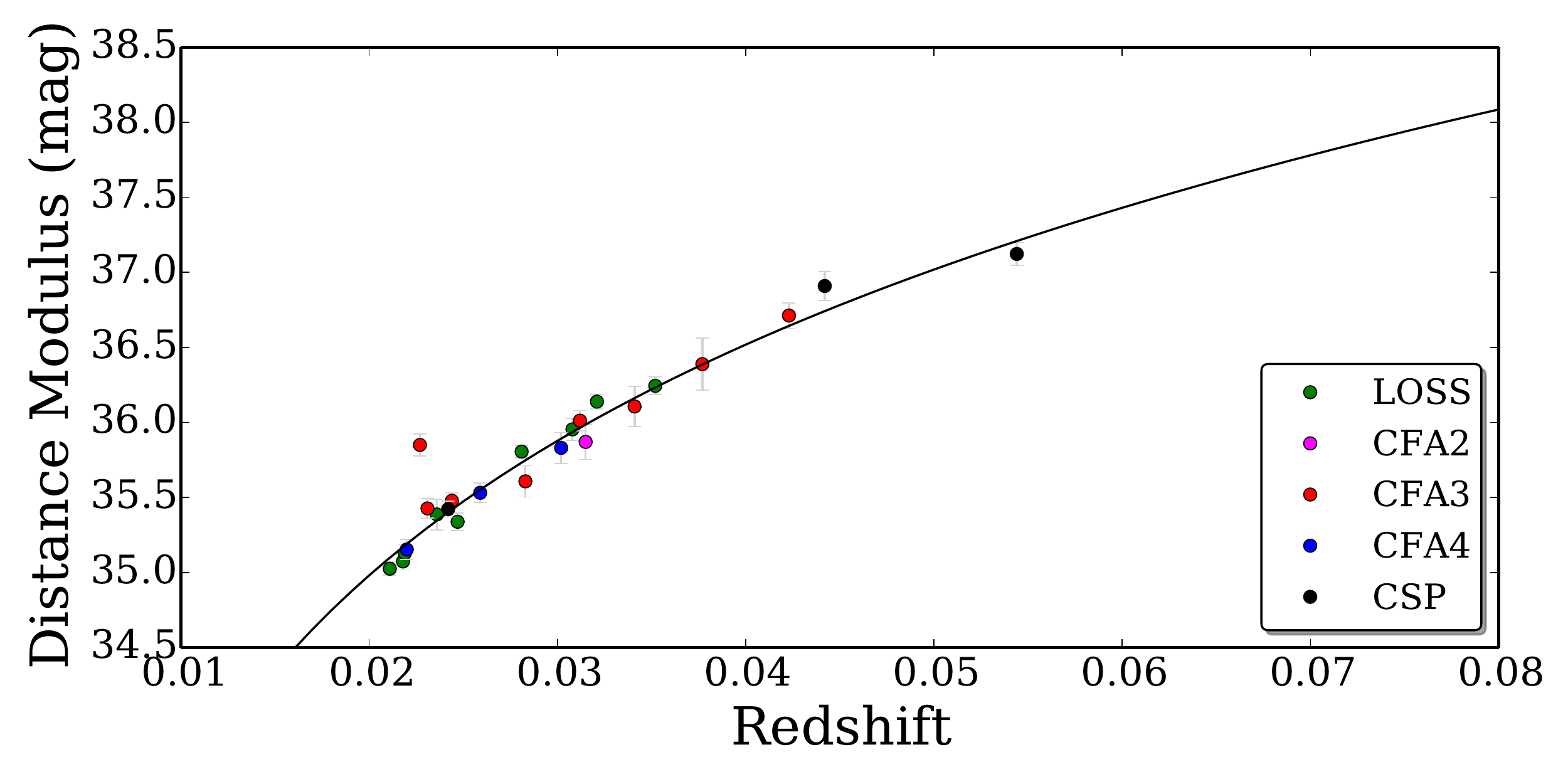}}
\caption{The SN~Ia Hubble redshift-distance relations. Upper panel shows distance modulus $\mu$ versus redshift $z$ for SNe~Ia. Lower panel shows the same relation for SNe where the aperture NUV surface brightnesses is brighter than \NUVregularCut\,mag\,arcsec$^{-2}$.
The color of each point shows the source of the SN light curve. In the lower panel, SN 2007bz is the single object with an outlying distance modulus ($0.60\pm0.07$\,mag).
}
\label{fig:hubblediagram}
\end{figure*}

Using images taken by the {\it GALEX} satellite, we measure the host-galaxy  surface brightnesses in the far- and near-ultraviolet (FUV and NUV) bandpasses within a circular $r=$ \aper\ aperture centered on the SN position.  %We show, in Figure~\ref{fig:NUV}, that 
SNe~Ia whose apertures have high NUV surface brightness (Fig.~\ref{fig:NUV}) exhibit an exceptionally small scatter among their Hubble residuals.
Among the \NUVMLCSeighteenHRtightCutsnnum\ SNe~Ia in environments brighter than \NUVtightCut\,mag\,arcsec$^{-2}$, the root-mean-square scatter in the Hubble residuals is \NUVMLCSeighteenHRtightCutall\,mag. When we examine only the \NUVLOWERRMLCSeighteenHRtightCutsnnum\ SNe~Ia with statistical uncertainty $\sigma_{\mu_{\rm SN}} < 0.075$\,mag for the distance modulus, the root-mean-square scatter is \NUVLOWERRMLCSeighteenHRtightCutall\,mag.
The only SNe we excluded when we computed the sample standard deviation is SN 2007bz, which exploded in a region of high surface brightness and has an $\sim8\sigma$ offset from the redshift-distance relation (HR $=0.60\pm0.07$\,mag).
Although our $R_V = 1.8$ MLCS2k2 fit to the light curve of SN 2007bz yields $A_V=0.26\pm0.07$\,mag, a published BAYESN fit instead favors a higher extinction of $A_V = 0.81\pm0.12$\,mag \cite{mandelnarayan11}, which would correspond to a significantly reduced Hubble residual.
The redshift-distance relation constructed using only SNe~Ia in environments with high NUV surface brightness exhibits significantly smaller scatter than that for the entire SNe~Ia sample (Fig.~\ref{fig:hubblediagram}).

To determine the statistical significance of finding a sample of SNe with Hubble-residual scatter $\sigma_{\rm HR}$, we performed 10,000 simulations in which we randomly shuffled the Hubble residuals of the parent SN sample, after removing outliers with $>0.3$\,mag.
For each shuffled sample of SNe, we next simulated the selection of a NUV surface-brightness upper limit that minimized the Hubble-residual scatter of the sample.   %To construct samples with similar sizes to the $N=$ \NUVMLCSeighteenHRtightCutsnnum\ analysis sample, 
Searching within $\pm0.2$\,mag of the \NUVtightCut\,mag NUV analysis upper limit in 0.05\,mag increments, we identify the upper limit that minimizes the shuffled sample's Hubble-residual scatter. 
The percentage of simulations that yielded a standard deviation smaller than $\sigma_{\rm HR}$ is the $p$ value. % statistic.
As shown in Table~\ref{tab:stddev}, only \NUVMLCSeighteenHRtightCutmc\ of simulated samples have a scatter smaller than the \NUVMLCSeighteenHRtightCutall\,mag that we measured for SNe~Ia in host environments with high UV surface brightness.

For the redshift distribution of the SNe that form the \NUVtightCut\,mag NUV sample, we used Monte Carlo simulations to calculate the expected contributions of peculiar velocities to Hubble-residual scatter. We computed  
$\sigma_{m}=$\,\NUVMLCSeighteenHRtightCutpeculiartwohundred\,mag for $\sigma_{\rm v}=$\,200 km s$^{-1}$, and
$\sigma_{m}=$\,\NUVMLCSeighteenHRtightCutpeculiarthreehundred\,mag for $\sigma_{\rm v}=$\,300 km s$^{-1}$.
Because the Hubble-residual scatter we measured for SNe Ia in host environments with high UV surface brightness is not much greater than that expected from peculiar motions alone, their intrinsic scatter in their luminosities after light-curve calibration is likely to be appreciably smaller than $\sim$0.08 mag ($\sim$4\% in distance).

\begin{figure*}
\centering
\subfigure{\includegraphics[angle=0,width=4.5in]{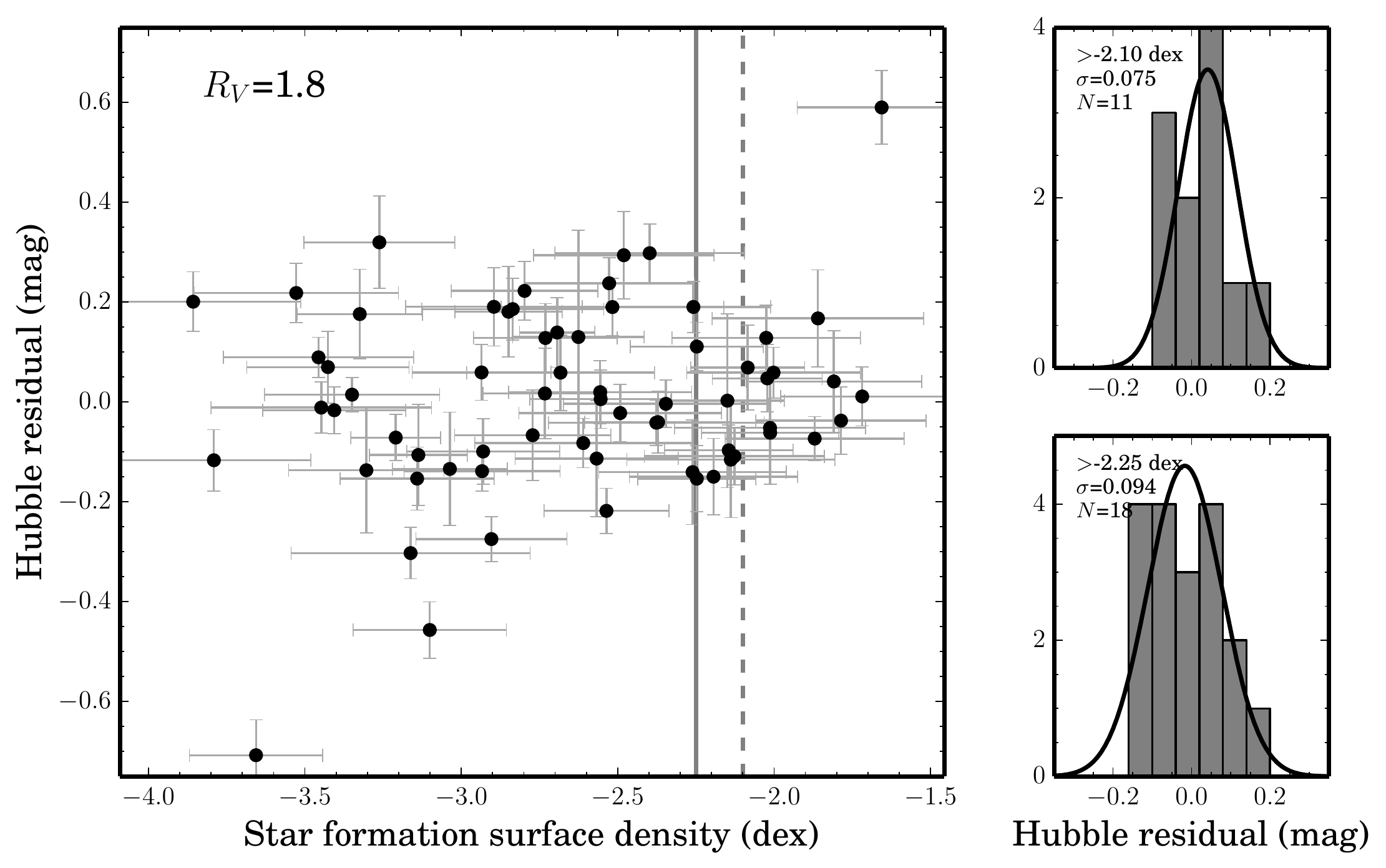}}
\caption{Hubble residuals against star-formation surface density within the $r=$ \aper\ aperture around SNe~Ia. 
Host-galaxy star-formation surface density measured within a circular $r=$ \aper\ aperture centered on the SN position.
Panels on the right show the distributions of Hubble residuals for SNe in regions with higher star-formation surface density than the \SFRAREADENregularCut\,dex limit marked by a solid vertical line and than the \SFRAREADENtightCut\,dex limit marked by the vertical dashed line.  As shown in Table~\ref{tab:sample}, a smaller number of host galaxies have the optical broadband photometry necessary to estimate the star-formation surface density within the $r=$ \aper\ circular aperture. 
}
\label{fig:SFRDEN}
\end{figure*}

We also estimated the average star-formation surface density [solar mass (M$_{\odot}$)\,yr$^{-1}$\,kpc$^{-2}$] within each circular $r=$ \aper\ aperture, when both optical ({\it ugriz} or {\it BVRI}) as well as FUV and NUV imaging of the host galaxy were available. 
The star-formation rate is computed by comparing the observed fluxes with predictions for stellar populations having a broad range of star-formation histories.
Although fewer host galaxies have the necessary imaging, Fig.~\ref{fig:SFRDEN} shows that the SNe having high star-formation density environments also have comparably small scatter in their Hubble residuals. The \SFRAREADENMLCSeighteenHRtightCutsnnum\ SNe~Ia with average star-formation surface density values in their apertures greater than \SFRAREADENtightCut\,dex exhibit $\sigma_{\rm HR}=$ \SFRAREADENMLCSeighteenHRtightCutall\,mag. Among randomly shuffled samples, only \SFRAREADENMLCSeighteenHRtightCutmc\ have a smaller standard deviation among their Hubble residuals, after searching within $\pm0.1$\,dex of the \SFRAREADENtightCut\,dex limit.

The 10\,kpc diameter of the host-galaxy aperture subtends an angle of $1.6''$ at $z=0.5$ and $1.3''$ at $z=1$. 
Therefore, for future cosmological analyses, the NUV surface brightnesses of high-redshift SN~Ia hosts within a circular $r=$ \aper\ aperture can be
measured from the ground in conditions with sub-arcsecond seeing, making possible precise measurements
of distances to high redshift.

The large UV surface brightnesses and star-formation densities of the environments of highly standardizable SNe~Ia, as well as the star formation evident from Sloan Digital Sky Survey (SDSS) images, reveal the existence of young, massive stars.  A reasonable conclusion is that the delay between the birth of the SN precursor and its explosion as a white dwarf is comparatively short.

The SDSS composite images in Fig.~\ref{fig:mosaic} show that the $r=$ \aper\ apertures include stellar populations of multiple ages, and the younger stellar populations are expected to dominate the measured NUV flux. 
While O-type and early B-type stars of masses $\gtrsim15\,{\rm M}_{\odot}$ are required to produce the ionizing radiation responsible for \ionpat{H}{ii} regions, stars with masses of $\gtrsim5\,{\rm M}_{\odot}$ having lifetimes of $\lesssim100$\,million years are responsible for the UV luminosity \cite{stewartfanelli00, gogartendalcanton09}.  A delay time of $\sim500$\,million years would be required for a SN~Ia progenitor to travel $\sim5$\,kpc with a natal velocity of 10\,km\,s$^{-1}$.

The total mass, metallicity, central density, and carbon-to-oxygen ratio of the white dwarf, as well as the properties of the binary companion, probably vary within the progenitor population of SNe~Ia. In theoretical simulations of both single-degenerate \cite{woosleykasen07,kasenropke09,simseitenzahl13} and double-degenerate \cite{kushnirkatz13,mollraskin14} explosions, the variation of many of these parameters can yield a correlation between light-curve decline rate and luminosity, but the normalization and slope of these predicted correlations generally show significant differences. Therefore, it is likely that variations in one, or possibly two, progenitor properties contribute significantly to the light-curve width/color/luminosity relation of the highly standardizable SNe~Ia population. The asymmetry of the explosion, which is thought to increase random scatter around the light-curve width/color/luminosity relation, may be small within this population and may possibly indicate that most burning occurs during the detonation phase \cite{kasenropke09}. A reasonable possibility is that the relatively young ages of the progenitor population correspond to a population with a smaller dispersion in their ages, leading to more uniform calibration. 

As we show in Table~\ref{tab:stddev}, for both the full sample and the SNe found in UV-bright environments, MLCS2k2 distances computed with $R_V=1.8$ yield a smaller Hubble-residual scatter than those computed with $R_V=3.1$.   %Improved Hubble-residual scatter for lower values of $R_V$ has also been observed for 
Although the low apparent value of $R_V$ may result in part from color variation unconnected to SN brightness \cite{scolnicriess14}, polarization data suggest that dust properties may also be important.
For a handful of well-sampled SNe~Ia where the extinction is large ($E(B-V)\gtrsim0.4$\,mag), the small intrinsic continuum polarization ($\lesssim0.3$\%) of SNe~Ia \cite{wangwheeler08} allows constraints on the wavelength-dependent polarization introduced by intervening dust \cite{patattaubenberger14}. Analyses of SN 1986G, SN 2006X, SN 2008fp, and SN 2014J show evidence for low values of $R_V$ and blue polarization peaks, consistent with a small grain size distribution \cite{patattaubenberger14}. In these cases, the polarization vector is aligned with the apparent local spiral arm structure, suggesting that the dust is interstellar rather than circumstellar.

A possibility is that SN environments exhibiting intense star formation may generate outflowing winds that entrain small dust grains, which might explain the evidence for low $R_V$ and the continuum polarization. Dust particles in the SN-driven superwind emerging from the nearby starburst galaxy M82 scatter light that originates in the star-forming disk, and the spectral energy distribution of the scattered light is consistent with a comparatively small grain size distribution \cite{huttonferreras14}.

\renewcommand\refname{References and Notes}

\printbibliography

\paragraph*{Acknowledgements} 
We thank S. Sim, J. C. Wheeler, J. Silverman, A. Conley, M. Graham, D. Kasen, I. Shivvers, and R. Kessler for useful discussions and comments on the paper, and J. Schwab for his help providing background about theoretical modeling. We are grateful to the staffs at Lick Observatory and Kitt Peak National Observatory (KPNO) for their assistance. The late Weidong Li was instrumental to the success of LOSS. A.V.F.'s supernova group at the University of California Berkeley has received generous financial assistance from the Christopher R. Redlich Fund, the TABASGO Foundation, and NSF grant AST-1211916. The Katzman Automatic Imaging Telescope and its ongoing operation were made possible by donations from Sun Microsystems, the Hewlett-Packard Company, AutoScope Corporation, Lick Observatory, NSF, the University of California, the Sylvia and Jim Katzman Foundation, and the TABASGO Foundation. The SLAC Department of Energy contract number is DE-AC02-76SF00515.  {\it GALEX} data are available from \url{http://galex.stsci.edu/GR6/}, SDSS data may be obtained at \url{http://www.sdss.org}, the KPNO imaging is archived at \url{http://portal-nvo.noao.edu}, and the Lick Observatory images are available from \url{http://astro.berkeley.edu/bait/public_html/iahostpaper/}.

\begin{table*}
\centering
\begin{tabular}{lcc}
\hline
Criterion & $\sigma_{\rm HR}$ ($R_V=1.8$)  & $\sigma_{\rm HR}$ ($R_V=3.1$) \\
\hline
UV and $\sigma_{\mu_{\rm SN}} < 0.075$ & \NUVLOWERRMLCSeighteenHRallall~($N$$=$\NUVLOWERRMLCSeighteenHRallsnnum) & \NUVLOWERRMLCSthirtyoneHRallall~($N$$=$\NUVLOWERRMLCSthirtyoneHRallsnnum)  \\
NUV SB $<$ \NUVtightCut & \NUVLOWERRMLCSeighteenHRtightCutall~($N$$=$\NUVLOWERRMLCSeighteenHRtightCutsnnum; $p$$=$\NUVLOWERRMLCSeighteenHRtightCutmc) & \NUVLOWERRMLCSthirtyoneHRtightCutall~($N$$=$\NUVLOWERRMLCSthirtyoneHRtightCutsnnum; $p$$=$\NUVLOWERRMLCSthirtyoneHRtightCutmc)  \\
NUV SB $<$ \NUVregularCut & \NUVLOWERRMLCSeighteenHRregularCutall~($N$$=$\NUVLOWERRMLCSeighteenHRregularCutsnnum; $p$$=$\NUVLOWERRMLCSeighteenHRregularCutmc) & \NUVLOWERRMLCSthirtyoneHRregularCutall~($N$$=$\FUVLOWERRMLCSthirtyoneHRregularCutsnnum; $p$$=$\FUVLOWERRMLCSthirtyoneHRregularCutmc)   \\
\hline
Full UV Sample & \NUVMLCSeighteenHRallall~($N$$=$\NUVMLCSeighteenHRallsnnum) & \NUVMLCSthirtyoneHRallall~($N$$=$\NUVMLCSthirtyoneHRallsnnum)  \\
NUV SB $<$ \NUVtightCut & \NUVMLCSeighteenHRtightCutall~($N$$=$\NUVMLCSeighteenHRtightCutsnnum; $p$$=$\NUVMLCSeighteenHRtightCutmc) & \NUVMLCSthirtyoneHRtightCutall~($N$$=$\NUVMLCSthirtyoneHRtightCutsnnum; $p$$=$\NUVMLCSthirtyoneHRtightCutmc)  \\
NUV SB $<$ \NUVregularCut & \NUVMLCSeighteenHRregularCutall~($N$$=$\NUVMLCSeighteenHRregularCutsnnum; $p$$=$\NUVMLCSeighteenHRregularCutmc) & \NUVMLCSthirtyoneHRregularCutall~($N$$=$\FUVMLCSthirtyoneHRregularCutsnnum; $p$$=$\FUVMLCSthirtyoneHRregularCutmc)   \\
\hline
Full SFR Sample & \SFRAREADENMLCSeighteenHRallall~($N$$=$\SFRAREADENMLCSeighteenHRallsnnum) & \SFRAREADENMLCSthirtyoneHRallall~($N$$=$\SFRAREADENMLCSthirtyoneHRallsnnum)  \\
$\Sigma_{\rm SFR} >$ \SFRAREADENtightCut & \SFRAREADENMLCSeighteenHRtightCutall~($N$$=$\SFRAREADENMLCSeighteenHRtightCutsnnum; $p$$=$\SFRAREADENMLCSeighteenHRtightCutmc) & \SFRAREADENMLCSthirtyoneHRtightCutall~($N$$=$\SFRAREADENMLCSthirtyoneHRtightCutsnnum; $p$$=$\SFRAREADENMLCSthirtyoneHRtightCutmc)  \\
$\Sigma_{\rm SFR} >$ \SFRAREADENregularCut & \SFRAREADENMLCSeighteenHRregularCutall~($N$$=$\SFRAREADENMLCSeighteenHRregularCutsnnum; $p$$=$\SFRAREADENMLCSeighteenHRregularCutmc) & \SFRAREADENMLCSthirtyoneHRregularCutall~($N$$=$\SFRAREADENMLCSthirtyoneHRregularCutsnnum; $p$$=$\SFRAREADENMLCSthirtyoneHRregularCutmc)   \\
\hline
\end{tabular}
\caption{The scatter among the SN~Ia Hubble residuals using $R_V=1.8$ and $R_V=3.1$. SB, surface brightness (mag arcsec$^{-2}$); $\Sigma_{\rm SFR}$, star formation surface density [solar mass ($M_{\odot}$) year$^{-1}$ kpc$^{-2}$]. We compute the standard deviation uncertainty through bootstrap resampling, after outliers with Hubble residuals of $>$\outlier\ were removed. A simulation found that the uncertainty of the standard deviation computed using bootstrap resampling is underestimated by $\sim$25\%, and we have corrected the estimates using this factor.  }
\label{tab:stddev}
\end{table*}

\renewcommand{\thetable}{S\arabic{table}}
\renewcommand{\thefigure}{S\arabic{figure}}
\setcounter{figure}{0} 
\setcounter{table}{0}

\clearpage
\begin{center}
\section*{\Large Supplementary Materials}
\end{center}

\paragraph{This PDF file includes the following:} % \mbox{} \\

\newenvironment{myitemize}
{ \begin{itemize}
    \setlength{\itemsep}{0pt}
    \setlength{\parskip}{0pt}
    \setlength{\parsep}{0pt}     }
{ \end{itemize}             }

\begin{myitemize}
\itemsep0em 
\setlength{\itemindent}{4.5pt}
\item[] Material and Methods
\item[] Supplementary Text
\item[] Figures S1 to S7
\item[] Tables S1 to S4
\item[] References 37--64
\end{myitemize}

\paragraph*{\large Materials and Methods}

\subparagraph{Selection of Light Curves.}
To include a SN in our sample, we require a photometric measurement before 5 days after {\it B}-band maximum, 
at least one photometric measurement after {\it B}-band maximum,
and at least three separate epochs having a signal-to-noise ratio (S/N) $>5$ measurement. Only photometry through 50 days 
after {\it B}-band maximum enters the fitting with MLCS2k2.

A fraction of the SNe in our sample were followed by multiple teams and published in more than one
of the light-curve collections listed in Table~\ref{tab:lcs}. 
We selected the light curve according to the order in Table~\ref{tab:lcs}, with light-curve collections at the beginning of the list having precedence. 
Rearranging the order does not significantly change the Hubble-residual statistics that we measure.  

\subparagraph{MLCS2k2 Light-Curve Fitting.}
Except for CSP SN~Ia photometry, for which we use magnitudes in the natural system, all fitting is performed to 
fluxes in the Landolt standard system \cite{landolt92}. Heliocentric SN~Ia redshifts are transformed to the cosmic microwave background (CMB) rest frame using the dipole velocity \cite{fixencheng96}. We correct fluxes for Milky Way extinction \cite{schlaflyfinkbeinerSFD11} using the O'Donnell 1994 reddening law \cite{odonnell94}. Since MLCS2k2 performs light-curve fitting to rest-frame fluxes, the observed SN fluxes are transformed, prior to fitting, from the observer frame into the rest frame using a K-correction \cite{nug02}. When performing MLCS2k2 fitting using SNANA, we apply an exponential prior on $A_V$ with $\tau=0.3$\,mag (PRIOR\_AVEXP $= 0.3$) convolved with a Gaussian kernel having $\sigma=0.02$\,mag (PRIOR\_AVRES $=$ 0.02). We obtain similar Hubble-residual scatter among SNe in UV-bright environments when priors instead with $\tau=0.1$\,mag and $\tau=0.5$\,mag are used while fitting.

To calculate the Hubble residuals of individual SNe, we first find the redshift-distance relation that minimizes the $\chi^2$ value of the fit to the MLCS2k2 distance moduli.  For the purpose of fitting for the redshift-distance relation, we add an uncertainty of 0.07\,mag in quadrature to the statistical uncertainty on each distance computed by MLCS2k2. After minimizing the $\chi^2$ value, we repeat the fit with only the SNe having Hubble residuals smaller than 0.3\,mag.   The best-fitting relation is not significantly affected by the specific values of the additional uncertainty, or of the criterion used to identify outlying SNe.

\begin{figure*}%[htp]
\centering
\subfigure{\includegraphics[angle=0,width=3.1in]{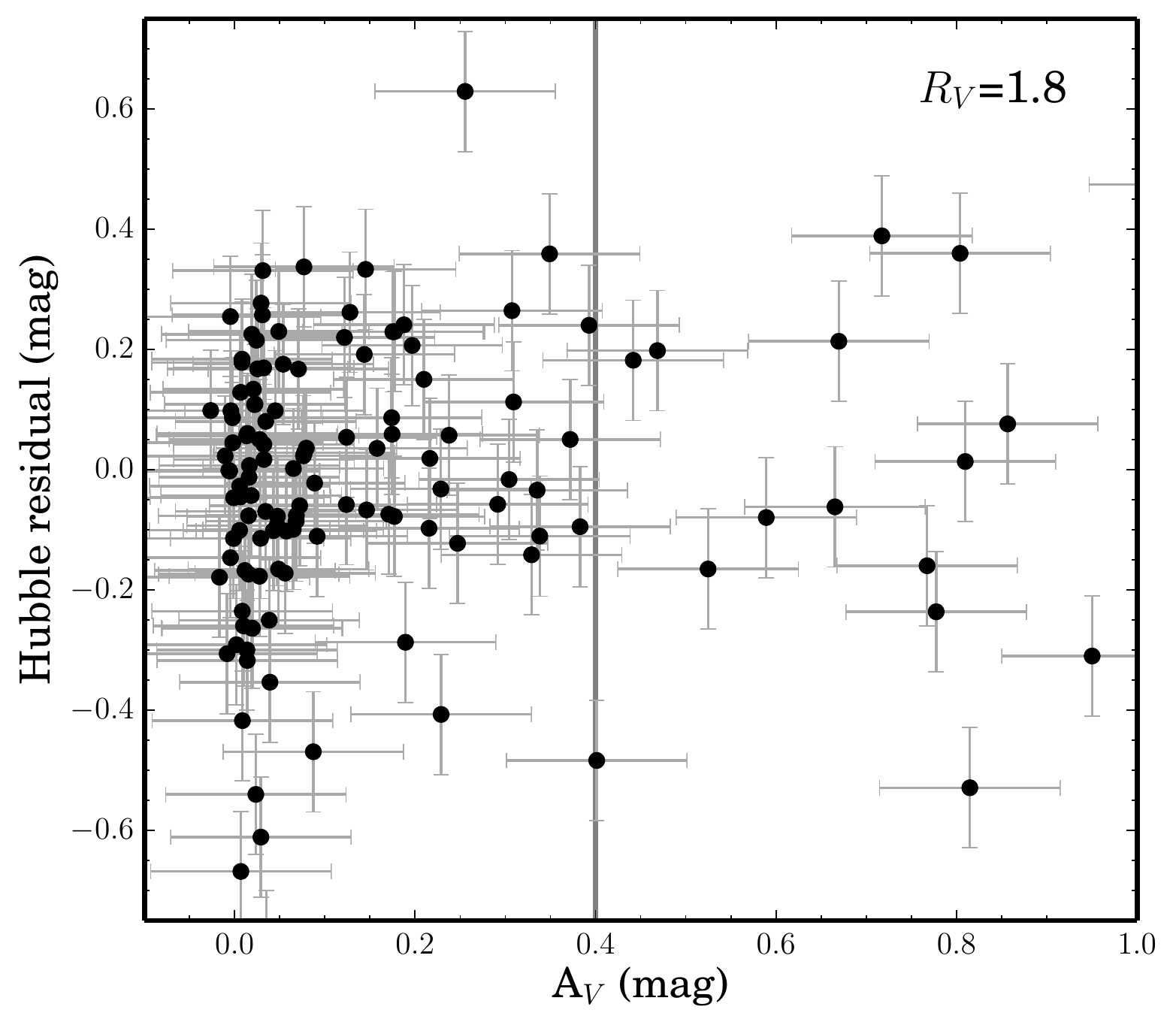}}
\subfigure{\includegraphics[angle=0,width=3.1in]{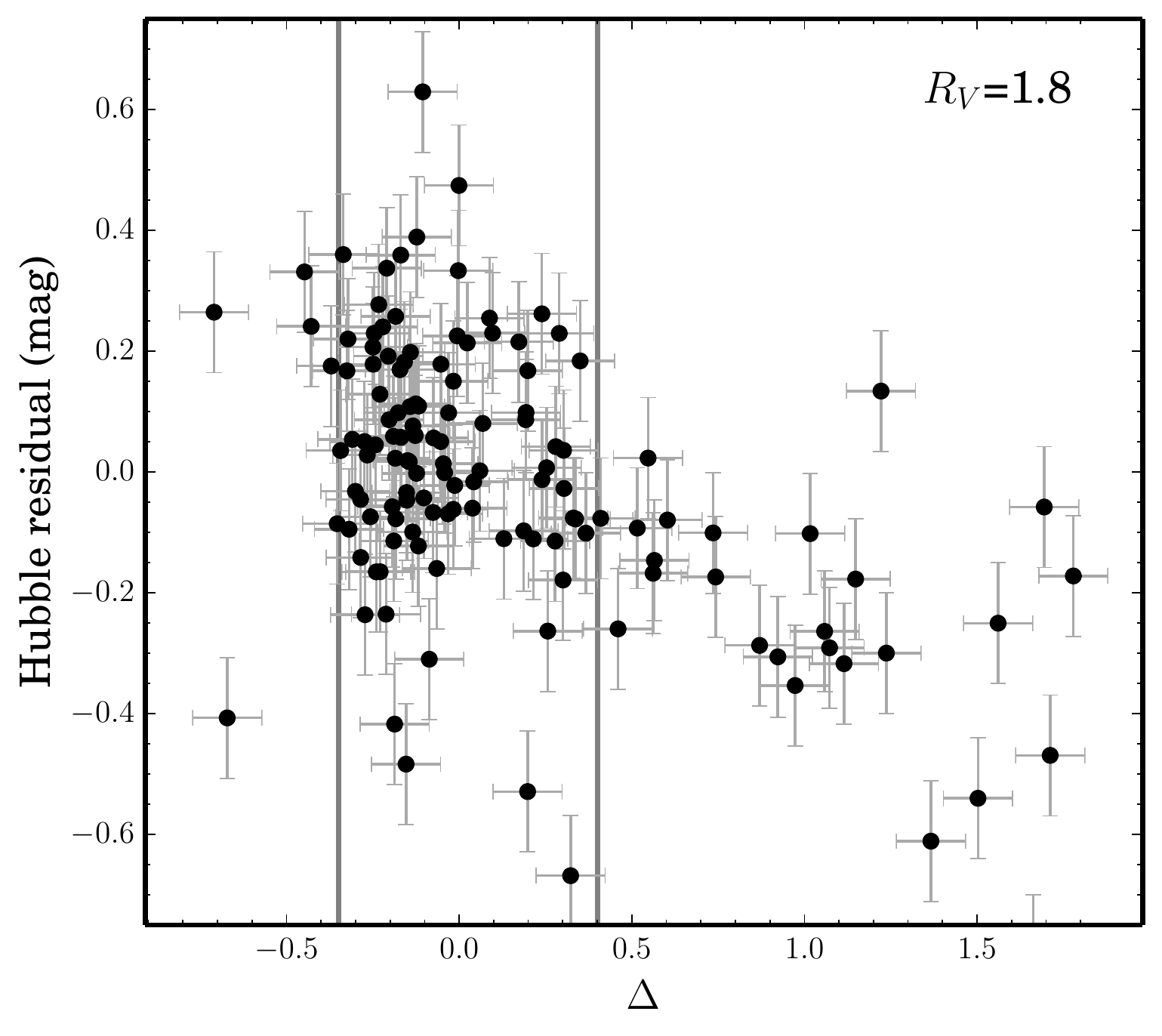}}
\caption{Hubble residuals plotted against MLCS2k2 $R_V=1.8$ light-curve parameters. When constructing our sample of SNe~Ia light curves, we apply the $A_V<$ \AVhigh\,mag upper limit shown in the left panel by the vertical line. We include only SNe with decline-rate parameter \deltalow\ $< \Delta <$ \deltahigh\ marked by the 
pair of vertical lines in the right panel. 
}
\label{fig:LCcuts}
\end{figure*}

\begin{figure*}%[htp]
\centering
\subfigure{\includegraphics[angle=0,width=3.1in]{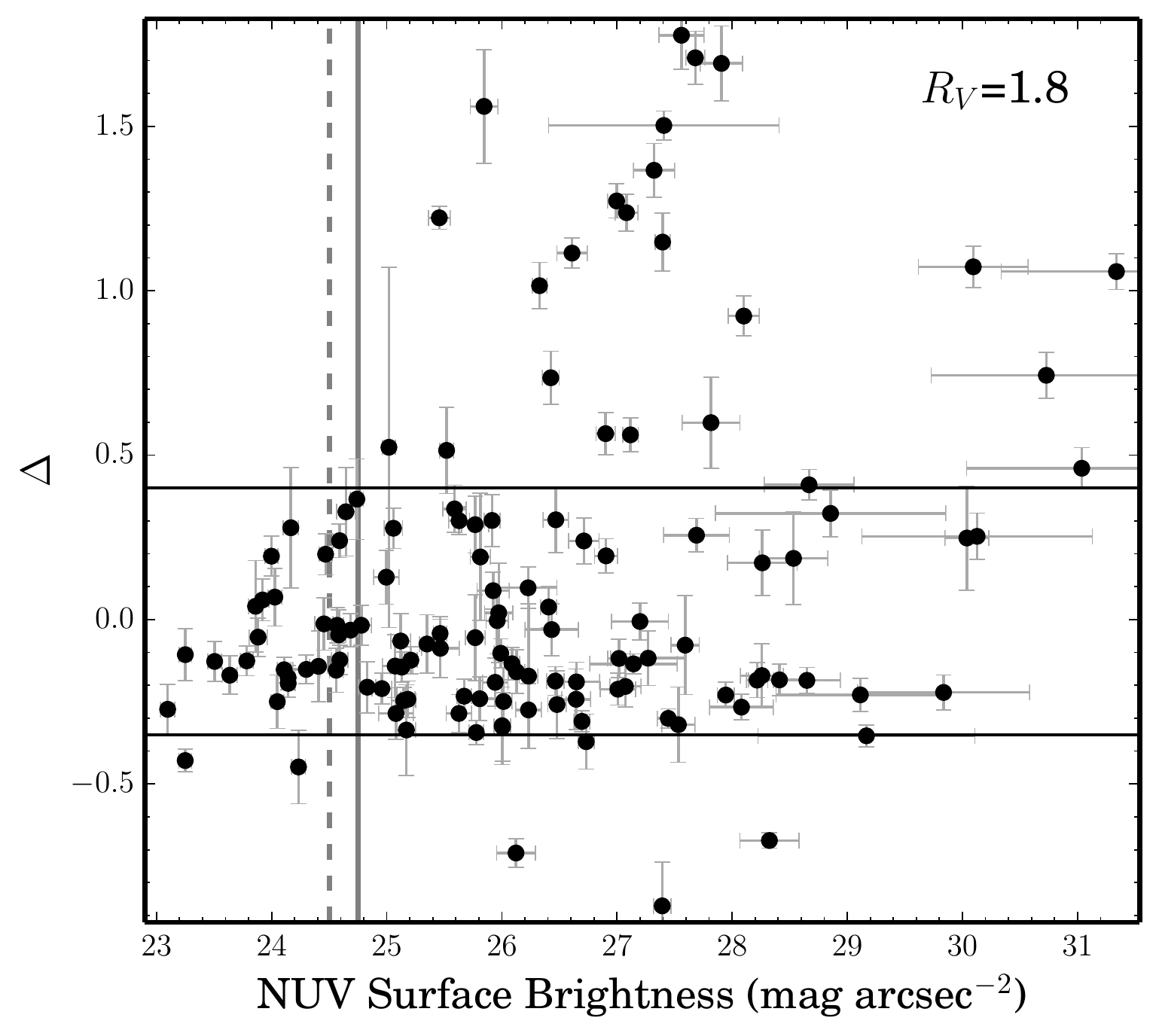}}
\caption{MLCS2k2 light-curve decline parameter $\Delta$ against mean NUV surface brightness within $r=$ \aper\ aperture around SN. 
More rapidly declining SNe~Ia have greater values of $\Delta$ and are, on average, less luminous explosions.
The parallel horizontal lines are the \deltalow~$< \Delta <$~\deltahigh~limits that we apply to construct our 
light-curve sample, while the parallel vertical lines show the \NUVtightCut\,mag\,arcsec$^{-2}$ (dashed) and \NUVregularCut\,mag\,arcsec$^{-2}$ (solid) NUV surface brightness upper limits. 
 }
\label{fig:DeltaVSUV}
\end{figure*}

Figures~\ref{fig:LCcuts} and \ref{fig:DeltaVSUV} show the light-curve parameter cuts that we apply to the sample of SN~Ia light curves. 
A principal purpose of light-curve cuts used in cosmological analyses is to remove SNe that may be systematically offset from the light-curve width/color/luminosity relation, that show extremely large dispersion around the redshift-distance relation, or that have light curves that indicate high reddening.  As Figs. \ref{fig:LCcuts}\ and \ref{fig:DeltaVSUV} show, we selected light-curve cuts that achieved these aims. 

As Figure~\ref{fig:DeltaVSUV} shows, fast-declining ($\Delta > 0.4$) SNe~Ia in our sample are absent from UV-bright host-galaxy environments.
Since these fast-declining SNe have Hubble residuals that are systematically low (Fig.~\ref{fig:LCcuts}), 
including these SNe in our analysis sample would increase the Hubble-residual scatter of the full SN~Ia sample, but not affect the UV-bright subset.
Therefore, the addition of the fast-declining SNe would increase the statistical significance of the 
comparatively small Hubble-residual scatter we observe among SNe in UV-bright regions. 
Nonetheless, we exclude the fast-declining SNe to show that SNe~Ia having typical light curves exhibit a smaller Hubble-residual scatter, when found in regions having high UV brightness.

\subparagraph{Calculating Statistical Significance.}
Through Monte Carlo simulations, we have assessed the probability that the small Hubble-residual scatter we measure for SNe~Ia in UV-bright
host-galaxy apertures is only a random effect. 
An additional possibility is that, from random effects, the SNe erupting instead, for example, in
UV-faint regions could have exhibited small apparent Hubble-residual
scatter. If we expect that we would have reported such an alternative
pattern as a discovery, then the $p$-values we compute should be
adjusted to account for multiple searches. 
For $n$ separate searches, the probability of finding at least one random pattern having significance $p$ is 
$p' = 1 - (1 - p)^n$.  
We have calculated, assuming a single search, a $p$-value of \NUVMLCSeighteenHRtightCutmc\ for the measured Hubble-residual scatter of \NUVMLCSeighteenHRtightCutall\,mag for SNe~Ia in regions with high NUV surface brightness (see Table~\ref{tab:stddev}).
Assuming instead $n = 4$ separate searches, for example, would yield an adjusted $p$-value of 0.8\%.

However, a principal focus of our analysis from its inception was to identify SNe~Ia in
UV-bright, star-forming environments using {\it GALEX} imaging, motivated by
a recent analysis of H$\alpha$ flux near explosion sites
\cite{rigaultcopin13}.
Indeed, while high UV surface brightness within the $r=$ \aper\
aperture provides strong evidence for a young stellar population,
interpretation of low average UV surface brightness is less
straightforward. As shown in Figure~\ref{fig:mosaic}, the apertures
with lower NUV surface brightness consist of a heterogeneous
mix of SNe in low-mass galaxies having small physical sizes, SNe with
large host offsets, and SNe erupting in old stellar populations.
Given the mixed set of environments, it probably would have been unlikely that we would
feel we had found a robust pattern. If SNe with very small scatter
exploded from the oldest stellar populations, we also note that such
an association probably would already have been identified using the
host-galaxy morphology, spectroscopy, or optical and infrared photometry \cite{hi09b,guptadandrea11,dandrea11,childressaldering13,johanssonthomas13,haydengupta13,pansullivan14}.

\subparagraph{KPNO Image Reduction and Calibration.}
We acquired {\it ugriz} imaging of SN~Ia host galaxies with the T2KA and T2KB cameras mounted on the Kitt Peak National Observatory (KPNO) 2.1\,m telescope in June 2009, March 2010, May 2010, October 2010, and December 2010. The raw images were processed using standard IRAF\footnote{http://iraf.noao.edu/} reduction routines. A master bias frame was created from the median stack of the bias exposures taken each night, and subtracted from the flat field and from images of the host galaxies. For each run, we constructed a median dome-flat exposure that we used to correct the host-galaxy images. To improve the flat-field correction, a median stack of all host-galaxy frames was computed after removing all  objects detected using SExtractor \cite{bert96}.

Even after dividing images by the flat field and by stacked, object-subtracted science images, the edge to the south of the T2KB CCD array showed a $>5$\% spatially dependent background variation. To remove this poorly corrected region of the detector, we trimmed the 400 pixel columns closest to south edge of the T2KB pixel array, after overscan subtraction. We constructed a fringe model for {\it z}-band images from object-subtracted science images, and scaled the model to best remove fringing from the affected images. 

\subparagraph{Lick Observatory Image Reduction and Calibration.} We used the bias and flat-field corrected {\it BVRI} ``template'' images of the host galaxies of SNe~Ia acquired after the SNe had faded, as part of the LOSS follow-up program \cite{fili01}.  The imaging was obtained using the 0.76\,m Katzman Automatic Imaging Telescope (KAIT) and the 1\,m Anna Nickel telescope at Lick Observatory \cite{ganeshalingam10}. During the period when the observations were taken, the CCD detector that was mounted on KAIT was replaced twice, while the {\it BVRI} filter set was replaced once. 

\subparagraph{Astrometric and Photometric Calibration of KPNO and Lick Observatory Imaging.}
We used the \textit{Astrometry.net} routine \cite{lang10}, which matches sources in each input image against positions in the USNO-B catalog,  to generate a World Coordinate System (WCS) for each KPNO\,2.1\,m, KAIT, and Nickel image. After updating each image with the WCS computed by \textit{Astrometry.net}, we performed a simultaneous astrometric fit that included every exposure of each host galaxy across all passbands using the {\tt SCAMP} package \cite{bertin06}, and the 2MASS \cite{skru06} point-source catalog as the reference.

The resulting WCS solution was passed to {\tt SWarp} \cite{bert02} and used to resample all images to a common pixel grid. To extract stellar magnitudes, we next used {\tt SExtractor} \cite{bert96} to measure magnitudes inside circular apertures with $4''$ radius. {\tt PSFEx} \cite{bertin11} was used to fit a Moffat point-spread function (PSF) model and calculate an aperture correction for each image. %Stars were selected through automatic 
We computed {\it ugriz} or {\it BVRI} zeropoints from the stellar locus using the publicly available {\tt big macs}\footnote{https://code.google.com/p/big-macs-calibrate/} package developed by the authors \cite{kellyvonderlinden14}.  The routine synthesizes the expected stellar locus for an instrument and detector combination using the wavelength-dependent transmission and a spectroscopic model of the SDSS stellar locus. 

\subparagraph{Host-Galaxy Photometry.}
To measure UV emission from host galaxies, we used both FUV and NUV imaging from the All-Sky Imaging Survey and Medium Imaging Survey performed by the {\it GALEX} \cite{martinfanson05} satellite. The {\it GALEX} FUV passband has a central wavelength of 1528\,\AA\ and a PSF full width at half-maximum intensity (FWHM) of $\sim6''$, while the central wavelength of the NUV passband is 2271\,\AA\ and the PSF FWHM is $\sim4.5''$.  Photometry at optical wavelengths was measured from {\it ugriz} images taken with the SDSS or the KPNO 2.1\,m telescope, or from {\it BVRI} images taken with KAIT or the 1\,m Nickel telescope at Lick Observatory. All optical images were convolved to have a PSF of $\sim4.5''$ to match that of the {\it GALEX} NUV images before extracting host-galaxy photometry.

When assembling mosaics of host-galaxy images, we record the MJD of each image. We exclude exposures taken from between two weeks prior to through 210 days past the date when the SN reached {\it B}-band maximum. 
For the SN host galaxies that have images without contaminating flux, we resample all exposures to a common grid centered on the SN explosion coordinates using the {\tt SWarp} software program.  We measure the host-galaxy flux within a circular $r=$ \aper\ aperture, and compute K-corrections using {\tt kcorrect} \cite{bl07}. 
If the uncertainty of a magnitude in a specific bandpass exceeds 0.3\,mag, then we use the magnitude synthesized from the {\tt kcorrect} model that best fits the full set of measured magnitudes, instead of adding a K-correction to the measured magnitude. 
We adjust surface-brightness estimates calculated from K-corrected magnitudes for $(1 + z)^{-4}$ cosmological dimming.
When both UV and optical host photometry are available, we estimate the star-formation surface density using the PEGASE2 \cite{fi99} stellar population synthesis models.  In Table~\ref{tab:data}, we list measured light-curve parameters, host-galaxy measurements, and Hubble residuals.

\paragraph*{\large Supplementary Text}

\begin{figure*}%[htp]
\centering
\includegraphics[angle=0,width=5in]{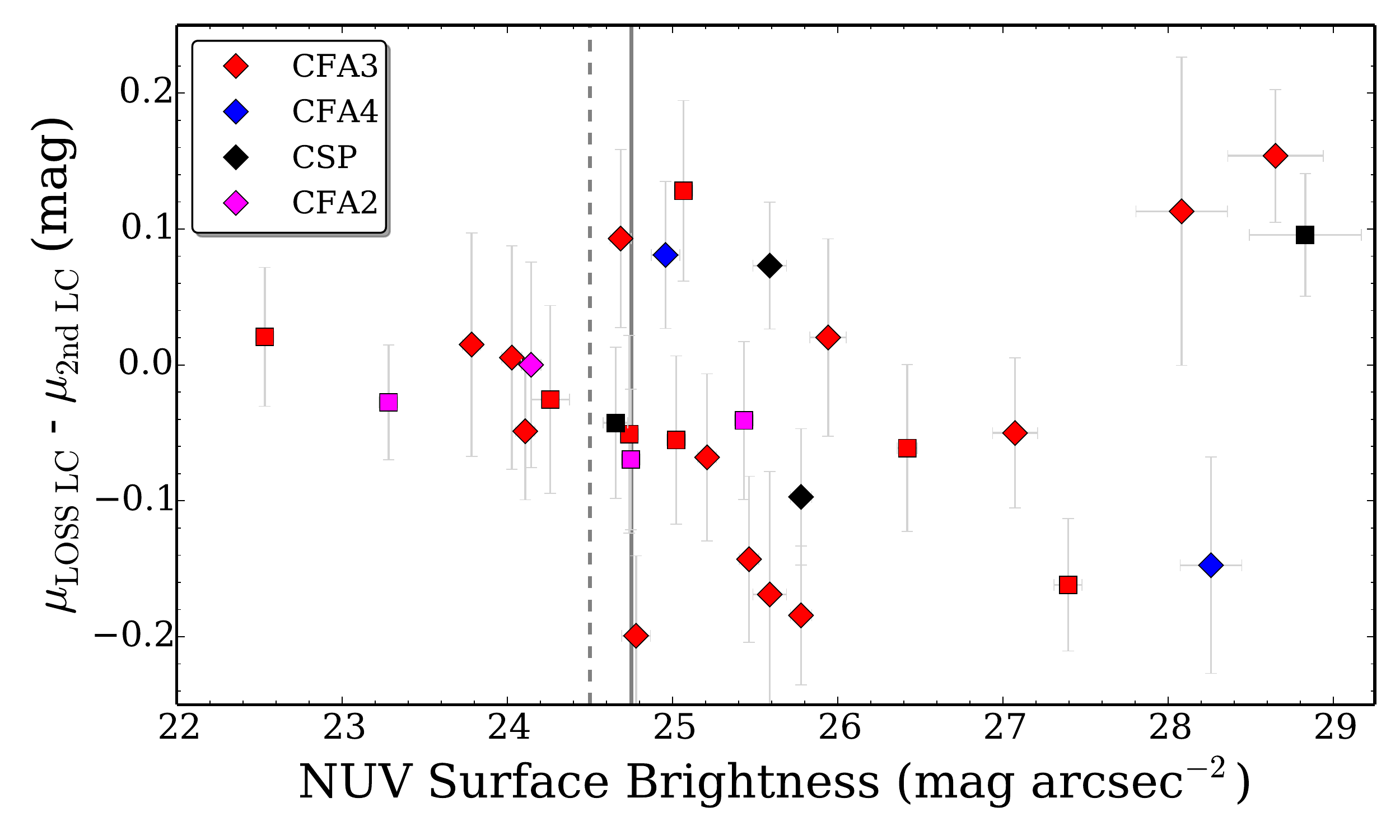}
\caption{Difference between MLCS2k2 distance moduli $\mu$ measured using LOSS and a second published light curve against mean NUV surface brightness within \aper\ of SN. 
SNe that erupt in regions with a high near-UV surface brightness may show greater agreement between distance measurements from independent light curves. 
Diamonds mark SNe having $z>0.02$ (included in the Hubble-residual sample), while squares mark SNe having $z<0.02$. 
The parallel vertical lines show the \NUVtightCut\,mag\,arcsec$^{-2}$ (dashed) and \NUVregularCut\,mag\,arcsec$^{-2}$~(solid) NUV surface brightness upper limits used to construct samples of SNe~Ia having low Hubble-residual scatter.
}
\label{fig:distComp}
\end{figure*}

\subparagraph{Additional Evidence for a Low-Scatter Population from SNe with Multiple Light Curves.}
A fraction of SNe have separate light-curve measurements published in the LOSS collection as well as in a second collection of light curves (CfA2, CfA3, CfA4, or CSP).
After completing an advanced draft of the paper, we compared the pairs of MLCS2k2 $R_V=1.8$ distance moduli to these SNe.
Since these distances are not affected by the peculiar motion of the host galaxies (unlike $\mu_z$), we can additionally study $z<0.02$ 
SNe which are not part of the Hubble-residual sample.  
Figure~\ref{fig:distComp} shows that there is greater agreement between the distance moduli 
to the SNe~Ia that explode in regions with high NUV surface brightness. 
This provides additional evidence, discovered after identifying the low-scatter population, that SNe~Ia in UV-bright regions yield more precise distances.

\begin{figure*}%[htp]
\centering
\subfigure{\includegraphics[angle=0,width=3.1in]{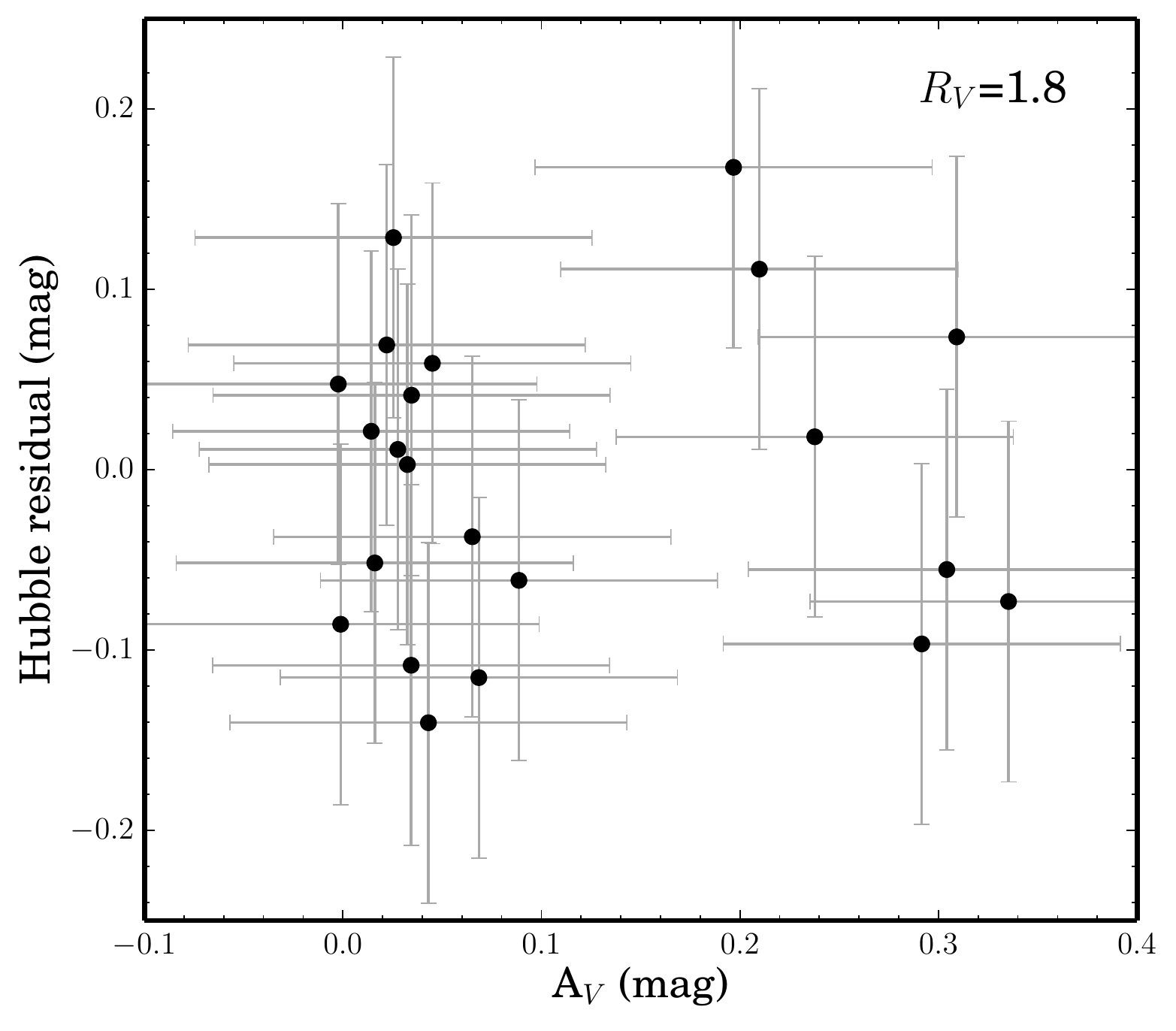}}
\subfigure{\includegraphics[angle=0,width=3.1in]{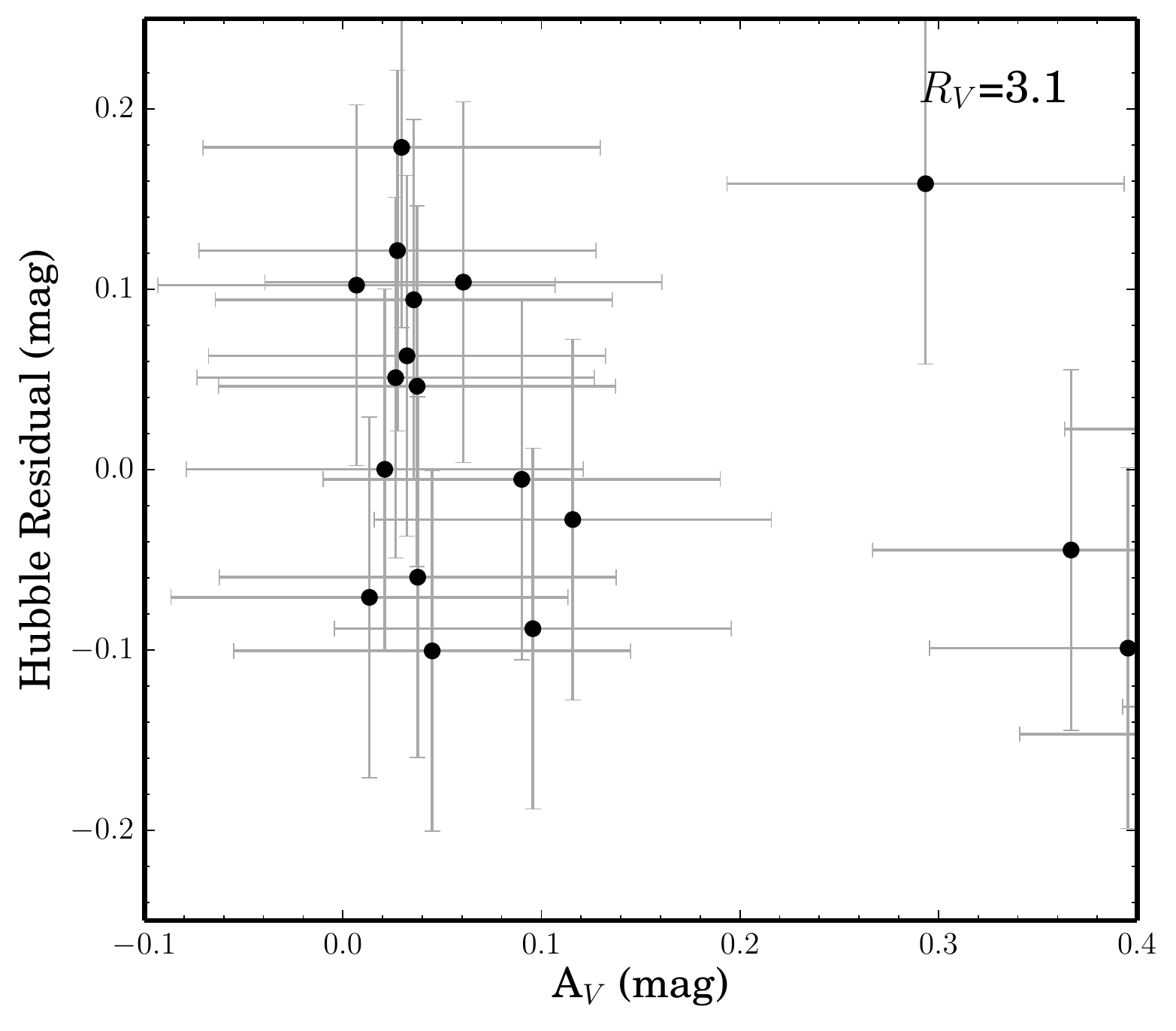}}
\caption{Hubble residuals against MLCS2k2 $A_V$ parameter for $R_V=1.8$ and $R_V=3.1$ model dust extinction laws for SNe with average NUV surface brightness $<$\NUVregularCut\,mag\,arcsec$^{-2}$  within $r=$ \aper\ aperture around SN. As shown in Table~\ref{tab:stddev}, distance moduli computed using MLCS2k2 with an $R_V=1.8$ dust extinction law exhibit a smaller dispersion among their Hubble residuals, than when computed using an $R_V=3.1$ model dust extinction law.  
} 
\label{fig:RV}
\end{figure*}

\subparagraph{$R_V$ and the Hubble Residuals of SNe in Regions with High NUV Surface Brightness.}
Figure~\ref{fig:RV} shows Hubble residuals against $A_V$ for MLCS2k2 distance moduli computed using $R_V=1.8$ and $R_V=3.1$.

\begin{figure*}%[htp]
\centering
\subfigure{\includegraphics[angle=0,width=4.5in]{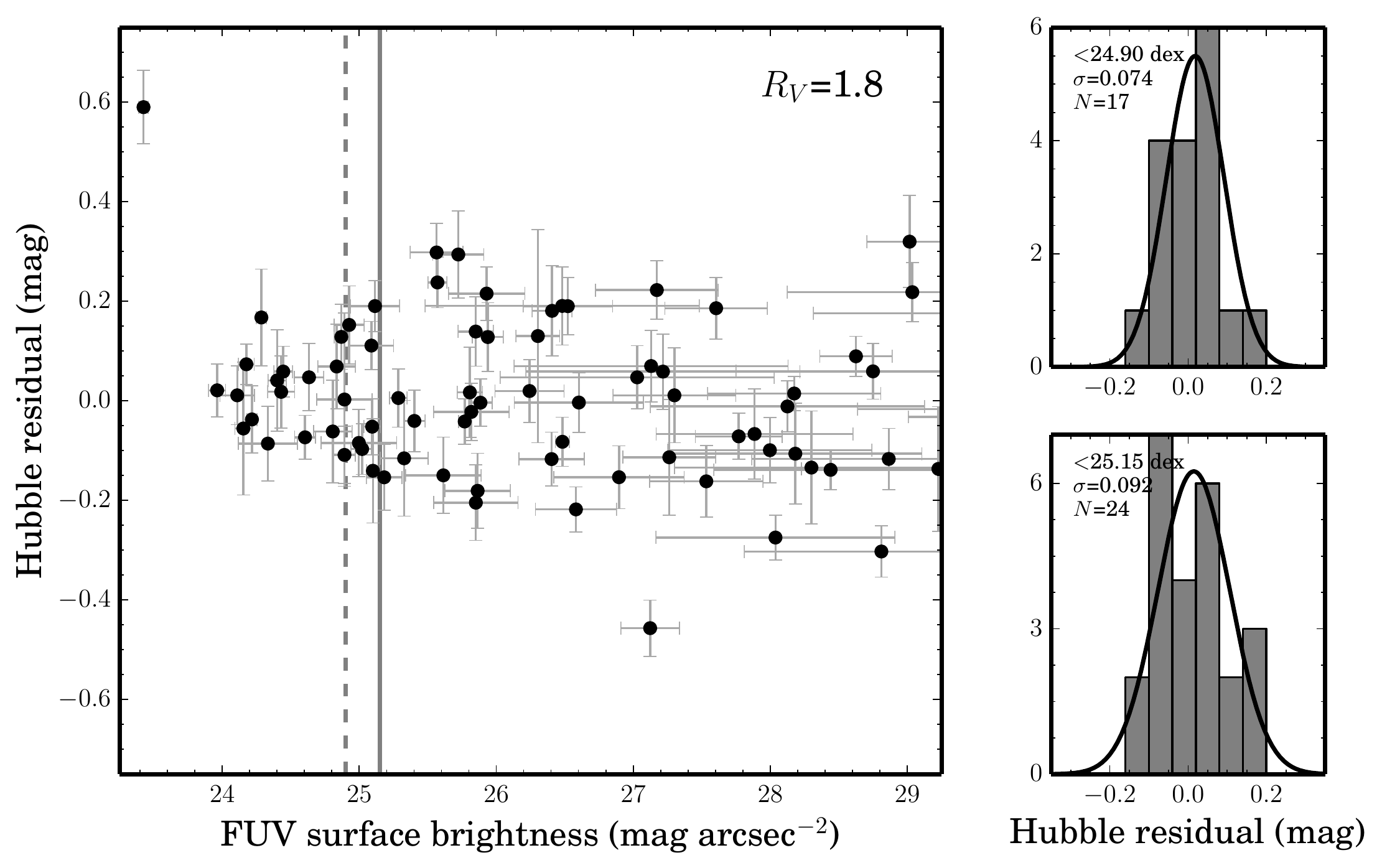}}
\caption{SNe~Ia in regions with high NUV (upper) and FUV (lower) surface brightness ($<$\NUVregularCut\,mag\,arcsec$^{-2}$ and \FUVregularCut\,mag\,arcsec$^{-2}$, respectively) inside a circular aperture with \aper\ radius exhibit smaller Hubble-residual dispersion. 
 }
\label{fig:FUV}
\end{figure*}

\subparagraph{FUV Surface Brightness.}
Figure \ref{fig:FUV} shows that the FUV surface brightness yields comparably small Hubble-residual scatter as NUV surface brightness.
Among the Hubble residuals of the $N=$ \FUVMLCSeighteenHRtightCutsnnum\ SNe~Ia in environments brighter than \FUVtightCut\,mag\,arcsec$^{-2}$, the root-mean-square scatter is \FUVMLCSeighteenHRtightCutall\,mag. When we examine only the $N=$ \FUVLOWERRMLCSeighteenHRtightCutsnnum\ SNe~Ia with distance modulus statistical uncertainty $\sigma_{\mu_{\rm SN}} < 0.075$\,mag, the root-mean-square scatter is \FUVLOWERRMLCSeighteenHRtightCutall\,mag.
Additional statistics are listed in Table~\ref{tab:stddev}.

\subparagraph{Color-Composite Images of Explosion Environments.}
Figure~\ref{fig:mosaic} displays panels of SDSS color-composite images of the host-galaxy environments for 
SNe~Ia whose circular $r=$ \aper\ apertures have high or low average NUV surface brightness.
These images show that SNe that occur within the host-galaxy outskirts, as well as in small low-mass galaxies, have 
low average NUV surface brightness within the aperture. 

\begin{figure*}%[htp]
\centering
\subfigure{\includegraphics[angle=0,width=4.5in]{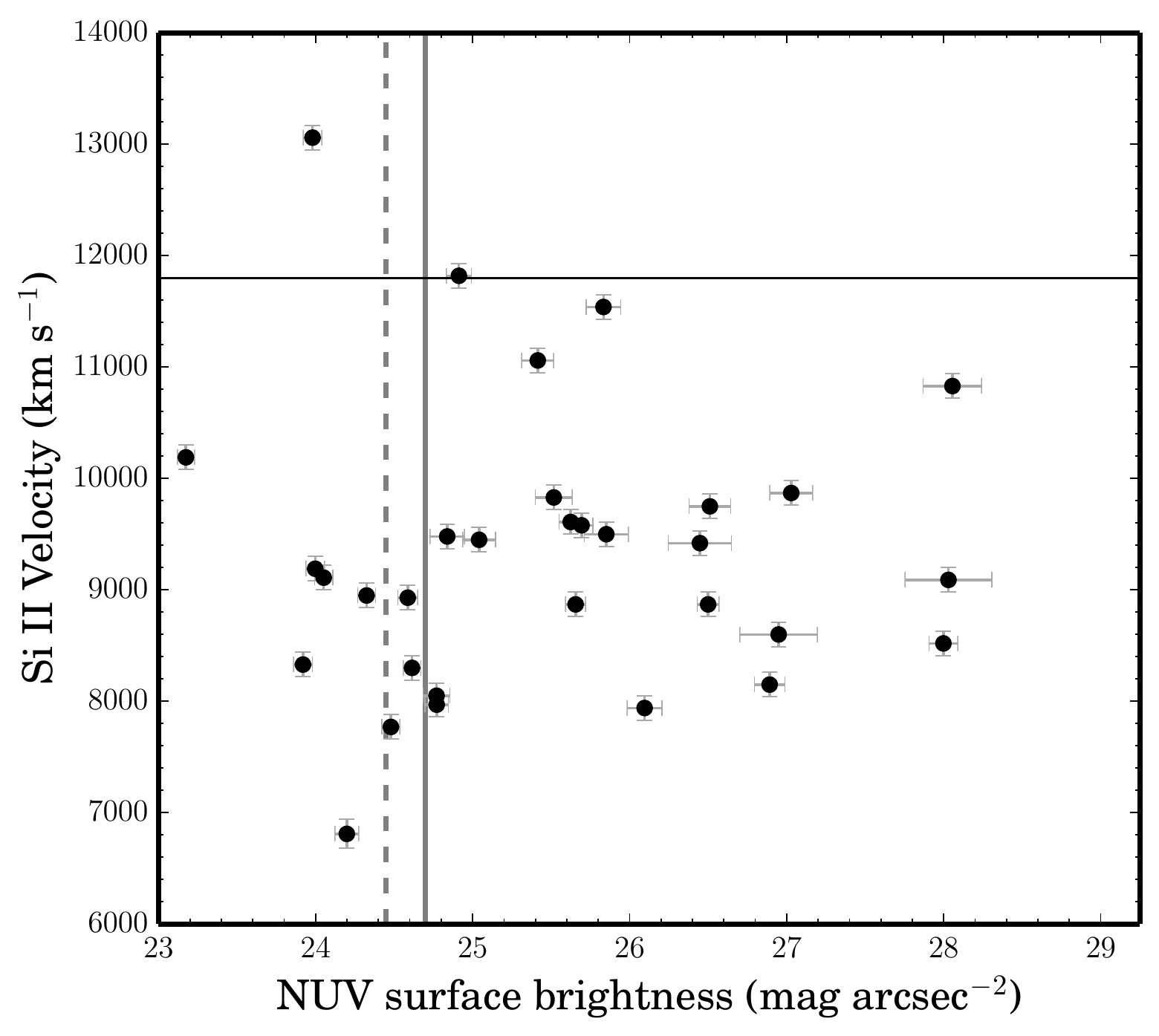}}
\caption{NUV surface brightness against \ionpat{Si}{ii} photospheric velocity of SNe~Ia. Horizontal line shows the 11,800\,km\,s$^{-1}$ line used by Wang et al. (2009) to separate low- and high-velocity SNe~Ia \cite{xwang09}. The population erupting in apertures with high NUV surface brightness does not differ strongly from that found in lower surface brightness regions.
 }
\label{fig:NUVsiiv}
\end{figure*}

\subparagraph{Si~II Velocity.}
In Figure~\ref{fig:NUVsiiv}, we show the velocities of SNe~Ia that have \ionpat{Si}{ii} measurements within seven days of 
{\it B}-band maximum brightness against NUV surface brightness within the circular $r=$ \aper\ aperture. The highly standardizable population
found in high surface brightness regions shows no strong differences in the distributions of \ionpat{Si}{ii} velocities
near maximum brightness.

\begin{table*}[htp!]
\centering
\begin{tabular}{ccc}
Light-Curve Dataset&Filters&References \\
\hline
Lick Observatory Supernova Search (LOSS) & {\it BVRI} & \cite{ganeshalingam10,leaman11,li11a} \\
CfA2 & {\it (U)BVRI} & \cite{jhakirshner06} \\
CfA3 & {\it (U)BVRIr'i'} & \cite{hi09a} \\
CfA4 & {\it (U)BVr'i'} & \cite{hickenchallis12} \\
Carnegie Supernova Project (CSP) & {\it (u)BVgri} & \cite{contrerashamuy10,stritzingerphillips11}  \\
\hline
\end{tabular}
\caption{SN~Ia light-curve samples used for the analysis. When fitting SN light curves, we did not include measured {\it U}- or {\it u}-band magnitudes. }
\label{tab:lcs}
\end{table*}

\label{sec:sample}
\begin{table*}
\centering
\begin{tabular}{lc}
\hline
Criterion & SNe \\
\hline
$N_{\rm DOF} > 5$ & 307\\
$\chi^2_{\nu} < 2$ & 292\\
$z>0.02$ & 165\\
$A_V < 0.4$\,mag & 143\\
$-0.35 < \Delta < 0.4$ & 107\\
MLCS2k2 & 107 \\
Not SNF20080522-000 & 106 \\
MW $A_V<0.5$\,mag & 103 \\
\hline
Uncontaminated FUV; NUV & 83 \\
$R_V=1.8$ Hubble residual $<0.3$\,mag & \NUVMLCSeighteenHRallsnnum \\ 
\hline
Uncontaminated FUV; NUV; {\it ugriz} or {\it BVRI}  & 67 \\
$R_V=1.8$ Hubble residual $<0.3$\,mag & \SFRAREADENMLCSeighteenHRallsnnum \\ 
\hline
\end{tabular}
\caption{Construction of SN sample. $\chi^2_{\nu}$ is the reduced chi-squared statistic calculated for the MLCS2k2 ($R_V=1.8$) light-curve fits. $A_V$ is the best-fitting extinction value computed by MLCS2k2. Milky Way extinction is computed from foreground dust maps \cite{schlaflyfinkbeinerSFD11}. We exclude host-galaxy images taken from 14 days prior to through 210 days after $B$-band maximum light.  }
\label{tab:sample}
\end{table*}

\begin{figure*}[t]
\centering
\subfigure{\includegraphics[angle=0,width=6.5in]{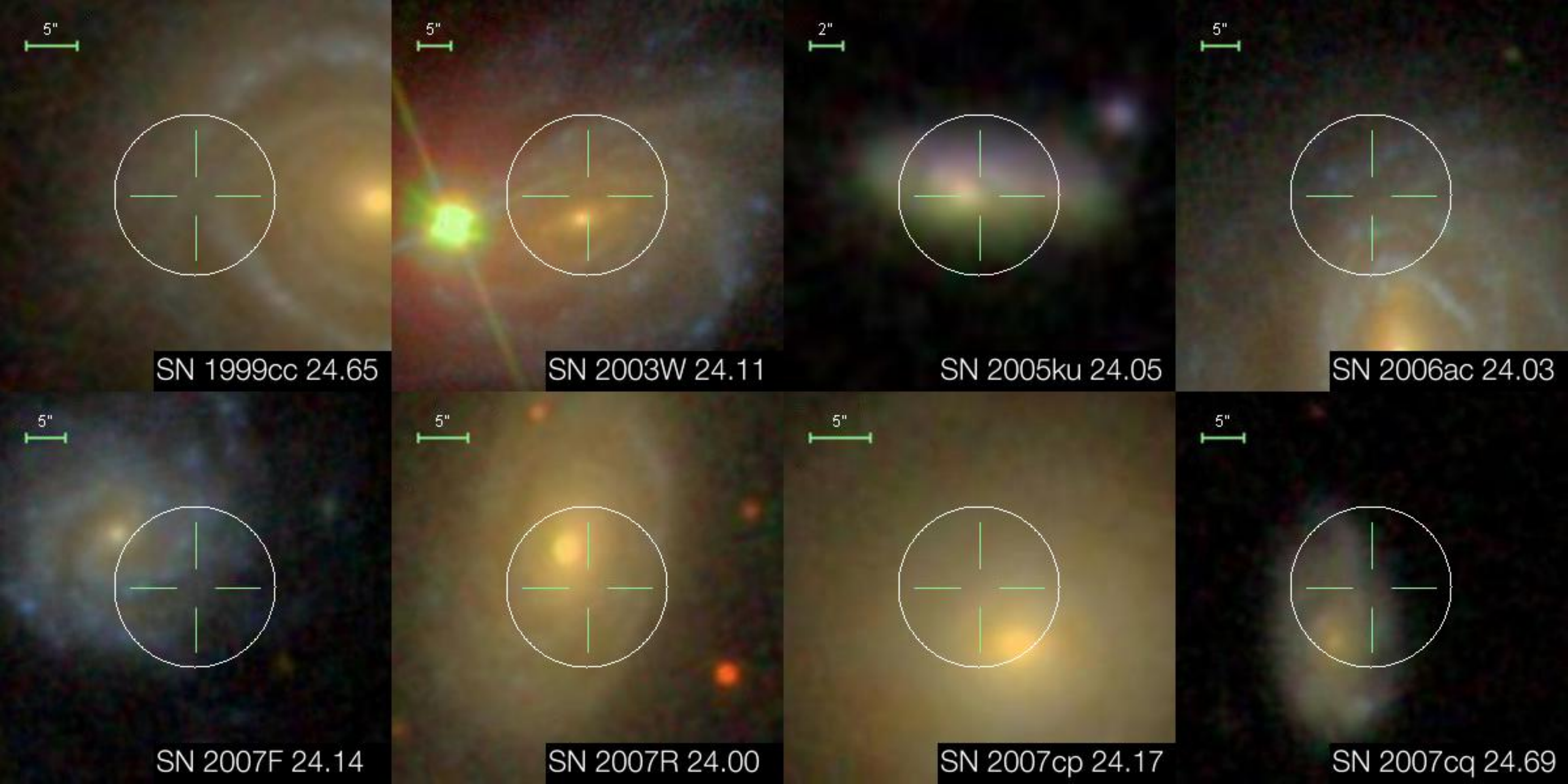}}
\subfigure{\includegraphics[angle=0,width=6.5in]{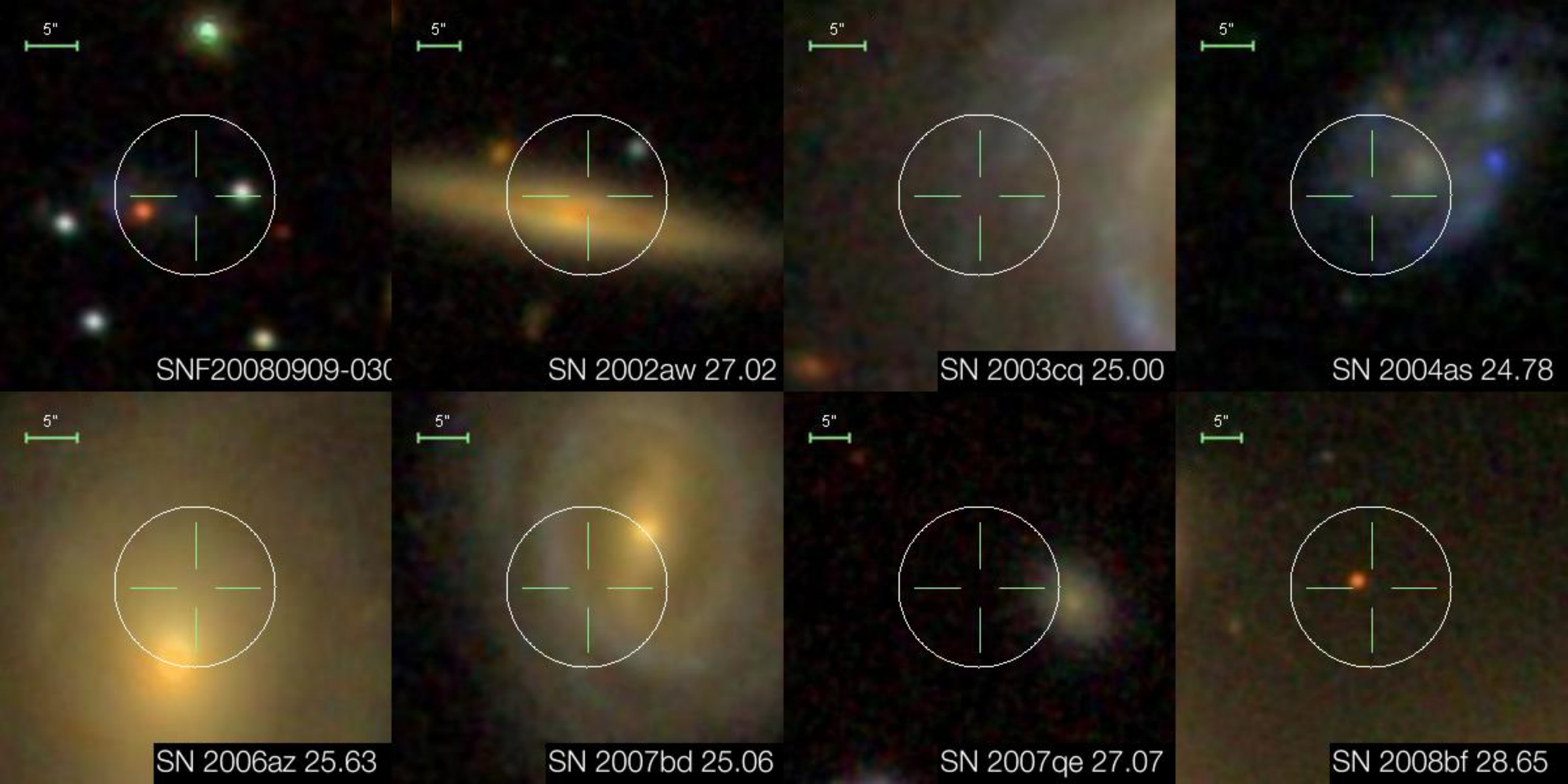}}
\caption{Representative SNe~Ia explosion sites with high and low average NUV surface brightness  within $r=$ \aper\ aperture around SN. White circles represent the apertures used for photometry. The upper mosaic shows SDSS color composite images of the host galaxies where the aperture NUV surface brightnesses is brighter than \NUVregularCut\,mag\,arcsec$^{-2}$, while the lower panels contain images of SNe with environments that have lower NUV surface brightness. Text next to each SN name gives the NUV surface brightness (in units of mag\,arcsec$^{-2}$), while the crosshairs are centered on the location of the explosion. }
\label{fig:mosaic}
\end{figure*}  

\clearpage
\begin{table*}[htp!]
\centering
\scriptsize
\caption{Host-Galaxy Environment}\begin{tabular}{lcccccc}
\hline
Name & $z_{\rm helio}$ & {\it FUV} SB & {\it NUV} SB & SFR Density & HR & HR\\
 &  & mag arcsec$^{-2}$ & mag arcsec$^{-2}$ & dex & $R_V$ = 1.8 & $R_V$ = 3.1\\
\hline
SN 1997dg & 0.034 & 25.72$\pm$0.19 & 25.82$\pm$0.10 & -2.48$\pm$0.29 & 0.29$\pm$0.09 & 0.04$\pm$0.11\\
SN 1999cc & 0.031 & 25.33$\pm$0.17 & 24.48$\pm$0.08 & -2.14$\pm$0.33 & -0.12$\pm$0.12 & -0.10$\pm$0.12\\
SN 1999dg & 0.022 & 27.89 (model) & 26.10$\pm$0.11 & -2.77$\pm$0.25 & -0.07$\pm$0.09 & -0.03$\pm$0.09\\
SN 2000dn & 0.032 & 27.61 (model) & 26.95$\pm$0.25 & -2.84$\pm$0.29 & 0.19$\pm$0.06 & 0.22$\pm$0.06\\
SN 2000fa & 0.021 & 25.02$\pm$0.08 & 24.00$\pm$0.06 & -2.15$\pm$0.21 & -0.10$\pm$0.05 & -0.13$\pm$0.06\\
SN 2001br & 0.021 & 27.17 (model) & 26.51$\pm$0.13 & -2.80$\pm$0.23 & 0.22$\pm$0.06 & 0.21$\pm$0.07\\
SN 2001cj & 0.024 & 28.63$\pm$0.26 & 27.84$\pm$0.07 & -3.46$\pm$0.30 & 0.09$\pm$0.04 & 0.13$\pm$0.04\\
SN 2001ck & 0.035 & 24.11$\pm$0.13 & 23.71$\pm$0.08 & -1.72$\pm$0.26 & 0.01$\pm$0.06 & 0.05$\pm$0.06\\
SN 2001cp & 0.022 & 28.13 (model) & 28.03$\pm$0.28 & -3.45$\pm$0.35 & -0.01$\pm$0.05 & 0.00$\pm$0.06\\
SN 2001ie & 0.031 & 29.23 (model) & 28.23$\pm$0.30 & -3.30$\pm$0.25 & -0.14$\pm$0.13 & -0.20$\pm$0.15\\
SN 2002aw & 0.026 & 27.53 (model) & 26.89$\pm$0.10 & \nodata & -0.16$\pm$0.07 & -0.24$\pm$0.09\\
SN 2002de & 0.028 & 24.18$\pm$0.06 & 23.67$\pm$0.06 & \nodata & 0.07$\pm$0.04 & 0.04$\pm$0.05\\
SN 2002eb & 0.028 & 28.17 (model) & 26.50$\pm$0.07 & -3.35$\pm$0.28 & 0.01$\pm$0.03 & -0.00$\pm$0.04\\
SN 2002el & 0.030 & 28.81 (model) & 27.51$\pm$0.29 & -3.16$\pm$0.38 & -0.30$\pm$0.05 & -0.27$\pm$0.05\\
SN 2002hu & 0.037 & 26.90 (model) & 26.45$\pm$0.20 & -3.14$\pm$0.25 & -0.15$\pm$0.06 & -0.12$\pm$0.07\\
SN 2003U & 0.028 & 25.10$\pm$0.05 & 24.61$\pm$0.06 & -2.26$\pm$0.30 & -0.14$\pm$0.10 & -0.12$\pm$0.10\\
SN 2003W & 0.020 & 24.60$\pm$0.08 & 23.98$\pm$0.06 & -1.87$\pm$0.29 & -0.07$\pm$0.04 & -0.11$\pm$0.05\\
SN 2003ch & 0.025 & 29.32 (model) & 28.15 (model) & -3.32$\pm$0.20 & 0.18$\pm$0.09 & 0.21$\pm$0.09\\
SN 2003cq & 0.033 & 25.61$\pm$0.28 & 24.84$\pm$0.11 & -2.19$\pm$0.27 & -0.15$\pm$0.08 & -0.20$\pm$0.09\\
SN 2003gn & 0.035 & 26.52 (model) & 25.52$\pm$0.12 & -2.52$\pm$0.36 & 0.19$\pm$0.06 & 0.14$\pm$0.06\\
SN 2003kc & 0.033 & 24.15$\pm$0.06 & 23.74$\pm$0.06 & \nodata & -0.06$\pm$0.13 & -0.11$\pm$0.20\\
SN 2004as & 0.031 & 25.09$\pm$0.16 & 24.65$\pm$0.09 & -2.25$\pm$0.21 & 0.11$\pm$0.05 & 0.05$\pm$0.05\\
SN 2004at & 0.022 & 26.48$\pm$0.05 & 25.89$\pm$0.06 & -2.61$\pm$0.35 & -0.08$\pm$0.05 & -0.04$\pm$0.05\\
SN 2004bg & 0.021 & 25.77$\pm$0.04 & 25.08$\pm$0.06 & -2.38$\pm$0.34 & -0.04$\pm$0.05 & -0.00$\pm$0.05\\
SN 2004br & 0.023 & 28.04 (model) & 26.64$\pm$0.15 & -2.90$\pm$0.24 & -0.27$\pm$0.04 & -0.23$\pm$0.05\\
SN 2004bw & 0.021 & 25.09$\pm$0.06 & 24.48$\pm$0.06 & -2.01$\pm$0.31 & -0.05$\pm$0.06 & -0.02$\pm$0.06\\
SN 2004ef & 0.031 & 26.40$\pm$0.24 & 25.41$\pm$0.10 & \nodata & -0.12$\pm$0.05 & -0.16$\pm$0.05\\
SN 2004gu & 0.045 & 27.77 (model) & 27.04$\pm$0.10 & -3.21$\pm$0.14 & -0.07$\pm$0.05 & -0.15$\pm$0.06\\
SN 2005ag & 0.079 & 28.44 (model) & 26.81 (model) & -2.93$\pm$0.25 & -0.14$\pm$0.04 & -0.18$\pm$0.05\\
SN 2005bg & 0.023 & 23.96$\pm$0.06 & 23.41$\pm$0.06 & \nodata & 0.02$\pm$0.05 & 0.04$\pm$0.06\\
SN 2005eq & 0.030 & 25.88$\pm$0.09 & 25.66$\pm$0.06 & -2.35$\pm$0.33 & -0.00$\pm$0.05 & -0.03$\pm$0.06\\
SN 2005eu & 0.035 & 25.40$\pm$0.08 & 25.32$\pm$0.06 & -2.37$\pm$0.24 & -0.04$\pm$0.06 & 0.00$\pm$0.06\\
SN 2005hc & 0.046 & 25.57$\pm$0.07 & 25.50$\pm$0.06 & -2.53$\pm$0.27 & 0.24$\pm$0.05 & 0.28$\pm$0.06\\
SN 2005iq & 0.034 & 25.93$\pm$0.28 & 25.85$\pm$0.14 & \nodata & 0.22$\pm$0.05 & 0.25$\pm$0.05\\
SN 2005ku & 0.050 & 24.29$\pm$0.04 & 23.78$\pm$0.06 & -1.86$\pm$0.34 & 0.17$\pm$0.10 & 0.14$\pm$0.13\\
SN 2005ms & 0.025 & 29.04 (model) & 28.00$\pm$0.09 & -3.53$\pm$0.33 & 0.22$\pm$0.06 & 0.25$\pm$0.07\\
SN 2006S & 0.032 & 25.11$\pm$0.18 & 25.04$\pm$0.11 & -2.26$\pm$0.25 & 0.19$\pm$0.05 & 0.16$\pm$0.07\\
SN 2006ac & 0.023 & 24.40$\pm$0.07 & 23.92$\pm$0.06 & -1.81$\pm$0.28 & 0.04$\pm$0.10 & 0.08$\pm$0.11\\
SN 2006an & 0.064 & 27.13 (model) & 27.00 (model) & -3.43$\pm$0.26 & 0.07$\pm$0.07 & 0.11$\pm$0.07\\
SN 2006az & 0.031 & 26.58$\pm$0.30 & 25.26$\pm$0.09 & -2.54$\pm$0.20 & -0.22$\pm$0.05 & -0.18$\pm$0.05\\
SN 2006cf & 0.042 & 24.84$\pm$0.14 & 24.20$\pm$0.08 & -2.08$\pm$0.18 & 0.07$\pm$0.09 & 0.11$\pm$0.09\\
SN 2006cj & 0.067 & 25.85$\pm$0.13 & 25.62$\pm$0.07 & -2.69$\pm$0.12 & 0.14$\pm$0.07 & 0.18$\pm$0.07\\
SN 2006cp & 0.022 & 26.24$\pm$0.25 & 25.83$\pm$0.11 & -2.56$\pm$0.29 & 0.02$\pm$0.06 & -0.01$\pm$0.07\\
SN 2006cq & 0.048 & 26.41$\pm$0.15 & 25.69$\pm$0.07 & -2.85$\pm$0.17 & 0.18$\pm$0.09 & 0.17$\pm$0.11\\
SN 2006en & 0.032 & 24.43$\pm$0.10 & 23.47$\pm$0.06 & \nodata & 0.02$\pm$0.07 & -0.06$\pm$0.09\\
SN 2006et & 0.022 & 27.30 (model) & 26.10$\pm$0.11 & \nodata & 0.01$\pm$0.10 & -0.06$\pm$0.12\\
SN 2006lu & 0.053 & 24.33$\pm$0.22 & 24.16$\pm$0.12 & \nodata & -0.09$\pm$0.07 & -0.09$\pm$0.08\\
SN 2006mp & 0.023 & 24.93$\pm$0.11 & 24.77$\pm$0.07 & \nodata & 0.15$\pm$0.08 & 0.12$\pm$0.10\\
SN 2006oa & 0.060 & 25.94$\pm$0.11 & 25.73$\pm$0.07 & -2.73$\pm$0.23 & 0.13$\pm$0.07 & 0.14$\pm$0.07\\
SN 2006on & 0.070 & 28.30 (model) & 26.51$\pm$0.14 & -3.04$\pm$0.18 & -0.13$\pm$0.11 & -0.21$\pm$0.15\\
SN 2006py & 0.060 & 28.18 (model) & 26.84$\pm$0.12 & -3.14$\pm$0.16 & -0.11$\pm$0.10 & -0.22$\pm$0.11\\
SN 2006sr & 0.024 & 24.87$\pm$0.05 & 24.32$\pm$0.06 & -2.03$\pm$0.30 & 0.13$\pm$0.07 & 0.16$\pm$0.07\\
SN 2007F & 0.024 & 24.45$\pm$0.07 & 24.05$\pm$0.06 & -2.00$\pm$0.28 & 0.06$\pm$0.05 & 0.09$\pm$0.06\\
SN 2007R & 0.031 & 24.63$\pm$0.11 & 23.80$\pm$0.06 & -2.02$\pm$0.18 & 0.05$\pm$0.07 & 0.09$\pm$0.07\\
SN 2007ae & 0.064 & 25.85 (model) & 25.59$\pm$0.14 & \nodata & -0.20$\pm$0.08 & -0.16$\pm$0.08\\
\hline
\end{tabular}
\end{table*}\begin{table*}[htp!]
\centering
\scriptsize\begin{tabular}{lcccccc}
\hline
Name & $z_{\rm helio}$ & {\it FUV} SB & {\it NUV} SB & SFR Density & HR & HR\\
 &  & mag arcsec$^{-2}$ & mag arcsec$^{-2}$ & dex & $R_V$ = 1.8 & $R_V$ = 3.1\\
\hline
SN 2007bd & 0.031 & 25.18$\pm$0.13 & 24.91$\pm$0.08 & -2.25$\pm$0.19 & -0.15$\pm$0.07 & -0.11$\pm$0.07\\
SN 2007bz & 0.022 & 23.43$\pm$0.03 & 23.17$\pm$0.05 & -1.66$\pm$0.27 & 0.59$\pm$0.07 & 0.52$\pm$0.10\\
SN 2007cp & 0.037 & 24.89$\pm$0.20 & 23.93$\pm$0.07 & -2.15$\pm$0.18 & 0.00$\pm$0.17 & 0.03$\pm$0.18\\
SN 2007cq & 0.026 & 24.89$\pm$0.08 & 24.59$\pm$0.06 & -2.13$\pm$0.29 & -0.11$\pm$0.06 & -0.08$\pm$0.06\\
SN 2007is & 0.030 & 24.81$\pm$0.14 & 24.28$\pm$0.08 & -2.01$\pm$0.22 & -0.06$\pm$0.10 & -0.04$\pm$0.13\\
SN 2007jg & 0.040 & 26.48 (model) & 26.02$\pm$0.25 & -2.90$\pm$0.28 & 0.19$\pm$0.08 & 0.22$\pm$0.09\\
SN 2007qe & 0.024 & 27.03 (model) & 27.03$\pm$0.14 & \nodata & 0.05$\pm$0.06 & 0.02$\pm$0.07\\
SN 2008Q & 0.008 & 31.12 (model) & 29.93$\pm$0.19 & -3.89$\pm$0.08 & -4.75$\pm$0.10 & -4.74$\pm$0.08\\
SN 2008Z & 0.021 & 29.02 (model) & 28.06$\pm$0.19 & -3.26$\pm$0.24 & 0.32$\pm$0.09 & 0.26$\pm$0.11\\
SN 2008ar & 0.026 & 25.56$\pm$0.19 & 24.77$\pm$0.09 & -2.40$\pm$0.30 & 0.30$\pm$0.06 & 0.32$\pm$0.07\\
SN 2008bf & 0.024 & 29.64 (model) & 28.42$\pm$0.29 & -3.41$\pm$0.23 & -0.02$\pm$0.05 & 0.02$\pm$0.05\\
SN 2008bz & 0.060 & 27.22 (model) & 26.04$\pm$0.23 & -2.68$\pm$0.30 & 0.06$\pm$0.08 & 0.09$\pm$0.08\\
SN 2008cf & 0.046 & 25.00$\pm$0.28 & 25.00$\pm$0.15 & \nodata & -0.08$\pm$0.07 & -0.05$\pm$0.07\\
SN 2008dr & 0.041 & 26.60 (model) & 25.69$\pm$0.07 & \nodata & -0.00$\pm$0.06 & 0.01$\pm$0.07\\
SN 2008fr & 0.039 & 27.12$\pm$0.21 & 26.33$\pm$0.08 & -3.10$\pm$0.25 & -0.46$\pm$0.06 & -0.42$\pm$0.06\\
SN 2008gl & 0.034 & 30.01 (model) & 29.98 (model) & \nodata & -0.03$\pm$0.08 & 0.00$\pm$0.08\\
SN 2008hj & 0.038 & 25.81$\pm$0.09 & 25.20$\pm$0.06 & -2.73$\pm$0.20 & 0.02$\pm$0.09 & 0.05$\pm$0.09\\
SN 2009D & 0.025 & 28.86 (model) & 28.29$\pm$0.27 & -3.79$\pm$0.31 & -0.12$\pm$0.06 & -0.08$\pm$0.07\\
SN 2009ad & 0.028 & 25.82$\pm$0.28 & 24.90$\pm$0.10 & -2.49$\pm$0.32 & -0.02$\pm$0.06 & 0.01$\pm$0.06\\
SN 2009al & 0.022 & 31.67 (model) & 29.74 (model) & -3.86$\pm$0.34 & 0.20$\pm$0.06 & 0.10$\pm$0.07\\
SN 2009do & 0.040 & 28.00 (model) & 26.07$\pm$0.07 & -2.93$\pm$0.24 & -0.10$\pm$0.07 & -0.09$\pm$0.08\\
SN 2009lf & 0.046 & 30.79 (model) & 28.66 (model) & -3.66$\pm$0.21 & -0.71$\pm$0.07 & -0.67$\pm$0.07\\
SN 2009na & 0.021 & 24.22$\pm$0.04 & 23.83$\pm$0.05 & -1.79$\pm$0.27 & -0.04$\pm$0.07 & -0.02$\pm$0.08\\
SN 2010ag & 0.034 & 25.86$\pm$0.24 & 25.50$\pm$0.11 & \nodata & -0.18$\pm$0.08 & -0.23$\pm$0.10\\
SN 2010dt & 0.053 & 26.30$\pm$0.16 & 25.94$\pm$0.08 & -2.63$\pm$0.21 & 0.13$\pm$0.21 & 0.16$\pm$0.22\\
SNF20080514-002 & 0.022 & 28.75 (model) & 26.90$\pm$0.10 & -2.94$\pm$0.22 & 0.06$\pm$0.06 & 0.10$\pm$0.06\\
SNF20080522-011 & 0.038 & 25.29$\pm$0.06 & 25.18$\pm$0.06 & -2.55$\pm$0.23 & 0.01$\pm$0.06 & 0.04$\pm$0.06\\
SNF20080909-030 & 0.031 & 27.26 (model) & 26.48$\pm$0.08 & -2.57$\pm$0.26 & -0.11$\pm$0.12 & -0.15$\pm$0.13\\
\hline
\end{tabular}
\caption{The NUV and FUV surface brightnesses are measured within a circular $r$\ $=$\ \aper~aperture centered at the SN position. When the uncertainty of the measured flux exceeds 0.3\,mag, we use the magnitude synthesized from the {\tt kcorrect} \cite{bl07} model that best fits the full set of measured magnitudes, instead of the K-corrected measured magnitude. The ``HR'' columns show the Hubble residuals of each SN Ia from the best-fitting redshift-distance relation for $R_V=1.8$ and $R_V=3.1$ MLCS2k2 light-curve fits.}
\label{tab:data}
\end{table*}
\end{document}